\documentclass[aps]{revtex4}

\usepackage{ulem}
\usepackage{euscript}
\usepackage{epsfig}

\begin{document}
\thispagestyle{empty}
\centerline{
\large{\bf Erratum: Kinetic Theory of a Spin-1/2 Bose-Condensed Gas}}
\centerline{
\large{\bf [J. Low Temp. Phys. 133, 323 (2003)]}}
\centerline{T. Nikuni and J. E. Williams}

\bigskip

Equation (76) should be
\begin{eqnarray*}
m\frac{\partial v_{c\mu}}{\partial t}&=&
-\frac{\partial \varepsilon_c}{\partial x_{\mu}}+
\frac{\hbar}{2}\vec{\Omega}_c\cdot\frac{\partial \vec{M}_c}{\partial x_{\mu}}
+\frac{\hbar}{2}v_{c\nu}
\frac{\partial \vec{M}_c}{\partial x_{\nu}}
\cdot\left(\vec{M}_c\times\frac{\partial \vec{M}_c}{\partial x_{\mu}}\right) \cr
&&-\frac{\hbar^2}{4mn_c}\frac{\partial\vec{M}_c}{\partial x_{\mu}}
\cdot\frac{\partial}{\partial x_{\nu}}\left( n_c \frac{\partial\vec{M}_c}
{\partial x_{\nu}} \right)
+\frac{1}{2}\left(\vec{M}_c\times\frac{\partial\vec{M}_c}
{\partial x_{\mu}}\right)\cdot{\rm Tr}(\uuline{\vec{\sigma}}
~ \uuline{R}).
\end{eqnarray*}
Here $\mu$ and $\nu$ are Cartesian components and repeated subscripts
are summed.

\setcounter{page}{0}

\clearpage
\title{Kinetic Theory of a Spin-$1/2$ Bose-Condensed Gas}

\author{T. Nikuni}
\affiliation{Department of Physics,
Faculty of Science, Tokyo University of Science,
1-3 Kagurazaka, Shinjuku-ku, Tokyo 162-8601, Japan.}
\author{J. E. Williams}
\affiliation{Electron and Optical Physics Division, National
Institute of Standards and Technology, Gaithersburg, Maryland
20899-8410}

%

\newcommand{\boldsymbol}[1]{\mbox{\boldmath $#1$}}


\begin{abstract}
We derive a kinetic theory for a spin-1/2 Bose-condensed gas of two-level atoms at finite temperatures.
The condensate dynamics is described by a generalized Gross-Pitaevskii equation for the two-component
spinor order parameter, which includes the interaction with the uncondensed fraction.
The noncondensate atoms are described by a quantum kinetic equation, which is a generalization of the
spin kinetic equation for spin-polarized quantum gases to include couplings to the condensate degree of freedom.
The kinetic equation is used to derive hydrodynamic equations for the noncondensate spin density.
The condensate and noncondensate spins are coupled directly through the exchange mean field.
Collisions between the condensate and noncondensate atoms give rise to an additional contribution to the spin
diffusion relaxation rate.
In addition, they give rise to mutual relaxation of the condensate and noncondensate due to lack of
local equilibrium between the two components.

\end{abstract}

\maketitle

\section{Introduction}
In this paper we present a kinetic theory for a trapped spin-1/2 Bose-condensed gas describing 
nonequilibrium collective dynamics of the density and spin at finite temperatures. 
Our work has many points of contact with several strands of research conducted over the past few
decades. 
In a way, this article provides a conceptual bridge between the very active field of Bose-Einstein 
condensation (BEC) in dilute gases and earlier work done on spin-polarized gases in the 
1980's~\cite{Schwarzschild1984a,Bashkin1986a,Silvera1986a,Meyerovich1987a,Laloe1988a,Betts1989a},
in which the field of ultracold gases has its roots.
It is interesting that, while cooling down to BEC was a central goal of the spin-polarized hydrogen work,
spin waves in a nondegenerate gas became a major topic for both
experiment~\cite{Johnson1984a,Nacher1984a,Gully1984a,Tastevin1985a,Bigelow1989a} and theory
\cite{Bashkin1981a,Lhuillier1982b,Lhuillier1982a,Levy1984a,Meyerovich1985a,Jeon1988a,Ruckenstein1989a,Jeon1989a}. 
After BEC was eventually observed in 1995 in the alkali atoms
\cite{Anderson1995a,Bradley1995a,Davis1995b}, 
however, the physics of spin waves in the new breed of experiments had been largely
overlooked---until very recently, when spin waves were observed at JILA
\cite{Lewandowski2002a,Oktel2002b,Fuchs2002a,Williams2002a,McGuirk2002a,Nikuni2002a,Kuklov2002a,Bradley2002a,Fuchs2002b}.
So an obvious direction for theory is to extend the earlier kinetic theories that were developed 
for a dilute spin-1/2 gas above 
$T_c$~\cite{Bashkin1981a,Lhuillier1982b,Lhuillier1982a,Levy1984a,Meyerovich1985a,Jeon1988a,Ruckenstein1989a,Jeon1989a} 
into the Bose-condensed regime, which is the main thrust of our paper.
Alternatively, our work can be viewed as an extension of the recent work on finite temperature 
kinetic theory for a single-component Bose-condensed gas
\cite{Stoof1999a,Walser1999a,Zaremba1999a,Gardiner2001a}
to now take into account the spin degree of freedom of the atoms.

Before placing our work in a broader context, it is useful to first clarify the type of system we
have in mind for our kinetic theory.
Experiments at JILA
\cite{Lewandowski2002a,McGuirk2002a,Hall1998b,Matthews1999b,Matthews1999a,Matthews1999c,Harber2002a,Lewandowski2002b}
have explored various properties of a dilute Bose gas of 
two-level atoms confined in a magnetic trap at finite temperatures.
Using a two-photon coupling field,
${}^{87}$Rb atoms can be prepared in a superposition of the two hyperfine states
$(F=1,M_F=-1)$ and $(2,1)$. This two-level atom can be treated as a spin-1/2 system.
These states are particularly nice to work with because they have nearly the same
magnetic moment, and thus experience the same trapping potential (to first order in the
magnetic field gradient).
In ${}^{87}$Rb, spin-exchange losses in the magnetic trap are reduced by more than a
factor of 1000 compared to other alkalis due to a fortuitous near-degeneracy of the
singlet and triplet scattering phase shifts
\cite{Myatt1997a,Julienne1997a,Kokkelmans1997a,Burke1997a}.
Ramsey fringe spectroscopy of the hyperfine splitting shows that the internal coherence--or transverse
spin--can be preserved throughout the gas for over a second, which is much
longer than the collisional relaxation time \cite{Harber2002a,Harber2002b}.
In this scenario, the transverse spin polarization can be treated approximately as a 
conserved quantity, although for long enough times the spins will dephase due to 
inhomogeneities.
These properties make this system ideal to study as a model spin-1/2 Bose gas.

\subsection{Pre-BEC work on spin waves}
Collective spin behavior is a well known property of metals with ferromagnetic or
antiferromagnetic ordering \cite{Halperin1969a}.
The idea that spin waves could also propagate in a paramagnetic system, such as ${}^3$He or a
normal metal, was first pursued in 1957 by Silin~\cite{Silin1957b,Silin1957a},
who generalized the Landau Fermi liquid theory
to treat the effect of an external magnetic field.
By including the off-diagonal magnetization terms in  the single-particle distribution function,
Silin derived a collisionless spin kinetic equation and showed that transverse spin waves could
propagate in a finite magnetic field.
The first experimental verification of spin waves in a paramagnetic system was reported by
Schultz and Dunifer in 1967 by measuring the electron spin resonance spectrum of thin metallic
slabs \cite{Platzman1967a,Schultz1967a}.
Shortly thereafter, Leggett and Rice predicted that, as a consequence of the collective spin
behavior, the spin diffusion coefficient should have a peak as a function of temperature,
which could be measured  using an NMR spin echo technique \cite{Leggett1968a,Leggett1970a}.
This so-called ``Leggett-Rice effect" was then verified in 1972 by Corruccini {\it{et al.}}
\cite{Corruccini1972a} in ${}^3$He-${}^4$He mixtures and normal liquid ${}^3$He.
A more direct observation of spin waves in liquid ${}^3$He came much later when the frequency 
spectrum of standing spin wave modes was measured by Masuhara {\it{et al.}} in 1984
\cite{Masuhara1984a,Candela1986a}.

All of this work dealt with dense Fermi liquids, so it came somewhat as a surprise when two independent
predictions appeared in 1982 by Bashkin~\cite{Bashkin1981a} and Lhuillier and Lalo\"{e}
\cite{Lhuillier1982b,Lhuillier1982a} that transverse spin waves can also propagate in dilute 
nondegenerate Bose and Fermi gases.
A few years earlier, dilute atomic hydrogen gas had been stabilized against molecular recombination at a few hundred degrees millikelvin by polarizing the electronic spin in a high magnetic 
field~\cite{Stwalley1976a,Silvera1980b,Cline1980a},
which provided a perfect testing ground for these predictions.
In order for a dilute gas to exhibit collective spin behavior, the temperature has to be low enough that
the thermal deBroglie wavelength $\lambda_{\rm{th}}$ is larger than the effective range of
interaction $r_0$, i.e., $\lambda_{\rm{th}} > r_0$.
If this condition is satisfied, then identical-particle symmetrization of the scattering
wavefunction for two colliding atoms is required and this gives rise to the quantum exchange interaction
needed for collective spin behavior.
The dilute hydrogen gas certainly satisfied this criterion, but yet was still far from the quantum
degenerate regime since $n^{1/3}\lambda_{\rm{th}}\ll 1$, which means that quantum statistics do not
play a role in thermodynamic properties.
To help dispel this paradox, Lhuillier and Lalo\"{e} offered the following picture,
which emerged from their microscopic treatment \cite{Lhuillier1982b,Lhuillier1982a}:
When two colliding atoms scatter off each other, their individual spins precess about their net
spin due to identical-particle symmetrization, even if the interaction has no explicit spin
dependence.
The cumulative effect of these spin rotations in successive collisions throughout the gas is a
collective oscillation of the transverse spin density.
This ``identical spin rotation effect" manifests itself in the kinetic theory as a nonlinear mean
field term proportional to the local spin density $\vec{S}({\bf r},t)$,
which exerts a torque on the spin current $\vec{{\bf J}}({\bf r},t)$.
The sign of this term for a Fermi gas is opposite to that of a Bose gas.

The first experimental verification of collective spin oscillations in a dilute gas came in 1984 by
Johnson {\it {et al.}}, who measured the spin wave resonances in the NMR spectrum of dilute spin
polarized hydrogen gas in the hydrodynamic regime \cite{Johnson1984a,Levy1984a}.
Later on this work was extended by Bigelow {\it {et al.}} to study the system in the low-density
collisionless regime \cite{Bigelow1989a}.
Shortly after the first hydrogen experiment, similar studies were done by Nacher {\it {et al.}} 
with dilute spin polarized ${}^3$He gas~\cite{Nacher1984a,Tastevin1985a} and by 
Gully and Mullin~\cite{Gully1984a} with a solution of ${}^3$He in ${}^4$He, 
which can be described as a dilute nondegenerate gas of quasiparticles.

Later work on collective spin effects in paramagnetic systems focused on understanding the effects
of quantum degeneracy.
In 1988, Jeon and Mullin~\cite{Jeon1988a,Jeon1989a,Mullin1992a} and, independently, 
Ruckenstein and L\' evy~\cite{Ruckenstein1989a}, used the Kadanoff-Baym nonequilibrium Green's
function approach~\cite{Kadanoff1962a} to obtain the kinetic equations describing quantum-degenerate 
(but nonsuperfluid) spin-1/2 Bose and Fermi gases.
The focus of these studies was to confirm with a more rigorous theory an earlier prediction made by
Meyerovich in 1985~\cite{Meyerovich1985a} that in a degenerate Fermi system the spin-diffusion
relaxation time $\tau_\perp$ for transverse spin waves should differ from the longitudinal 
relaxation time $\tau_\parallel$.
Jeon and Mullin showed that, while the longitudinal relaxation time exhibits the expected divergent
behavior $\tau_\parallel\sim 1/T^2$, the transverse relaxation time saturates to a finite value as 
$T \rightarrow 0$.
In addition to depending on temperature, this anisotropy also depends on the degree of polarization
(which is why the earlier theories of Silin~\cite{Silin1957b,Silin1957a} and 
Leggett~\cite{Leggett1968a,Leggett1970a} overlooked it,
since they both take the limit of zero polarization in equilibrium).
Experimental evidence supporting this prediction appeared in the mid 1990's from 
Wei {\it {et al.}}~\cite{Wei1993a} and Ager {\it{et al.}}~\cite{Ager1994a}, 
who used NMR spin echo techniques to measure the spin-diffusion coefficients in spin polarized ${}^3$He.
Very recently, however, similar experiments were conducted by Vermeulen and Roni~\cite{Vermeulen2001a}
at a much higher polarization and they find no evidence for polarization induced spin wave damping
\cite{Fomin1997a}.
It is clear that further experiments at higher polarizations and lower temperatures could give
more definitive results.
Two-component degenerate Fermi gases of alkali atoms,
which are being actively studied by several different groups~\footnote{For a concise overview of work on
Fermi gases, see the review article by Anglin and Ketterle \cite{Anglin2002a}.},
could play a deciding role in answering this question, since polarizations of  $100\%$ 
are possible in those systems.

The effects of degeneracy in a spin-1/2 Bose gas were not explored in great detail before BEC was achieved in 1995, since experiments had not operated in this regime before that time. Sparked by the paper in 1976 by Stwalley and Nosanow~\cite{Stwalley1976a} that made a strong case  for BEC in spin-polarized hydrogen,
a number of theory papers appeared over the following several years on BEC in a dilute spin-1/2 gas
(see for example, Ho~\cite{Ho1982a} and references therein),
emphasizing that the internal degrees of freedom made the system potentially much richer than superfluid
${}^4$He, and would have a closer resemblance to superfluid ${}^3$He in terms of its superfluid dynamics
\cite{Ho1982a} and symmetry breaking properties\cite{Siggia1981a}.
The theory of spin waves in a Bose-condensed spin-1/2 gas, however,  was not pursued,
despite the fact that predictions for a classical gas by Bashkin~\cite{Bashkin1981a} and Lhuillier and
Lalo\"{e}~\cite{Lhuillier1982b,Lhuillier1982a} appeared around that time in the early 1980's.
Steps were certainly taken in this direction when Jeon and Mullin~\cite{Jeon1988a,Jeon1989a,Mullin1992a}
and  Ruckenstein and L\'evy~\cite{Ruckenstein1989a} extended the kinetic theory to lower temperatures
where quantum statistics play a role~\cite{Jeon1991a},
but spin waves in the Bose-condensed regime were so far never treated.

\subsection{Post-BEC work on spin waves}
After BEC was achieved in 1995 with magnetically trapped alkali atoms,
there was no immediate attempt to investigate spin waves in the Bose-condensed regime.
In general, it is very difficult to maintain a spinor gas in a magnetic trap,
thus the main focus of research was on single-component condensation,
in which only a single hyperfine state is occupied.
Unlike in the earlier hydrogen experiments, the alkali atoms are typically placed in a very weak
magnetic field (around a millitesla), and so the total hyperfine spin governs the magnetic confining 
potential.
There are only three low-field seeking states in ${}^{87}$Rb, ${}^{23}$Na, and ${}^{7}$Li: $(F=2,M_F=2)$, $(2,1)$, and $(1,-1)$, which can be magnetically trapped.
Conventional wisdom is that having a true spinor system in a magnetic trap is precluded by spin-exchange 
collisions, which transfer atoms to untrapped hyperfine states.
Typical loss rates due to this density-dependent process are on the order of $10^{-11} \rm{cm}^3/s$ for the alkali
atoms, which can limit the lifetime of multicomponent gases to less than a millisecond.
One way to circumvent this problem is to load the atoms into an all-optical trap,
as first done at MIT in 1998 with ${}^{23}$Na atoms prepared in the $F=1$ hyperfine 
multiplet~\cite{Stenger1998a,Stenger1998b},
which are simultaneously trapped in the same confining optical potential.
In this spin-1 system, spin-exchange collisions play a beneficial role by allowing the relative
populations between the spin states to equilibrate in order to reach the thermodynamic
ground state.

To have a spin-1/2 Bose gas, we need two hyperfine states of atoms in a gas.
In Appendix~\ref{two-level}, we elaborate on several two-level systems at our disposal
in dilute atomic gases. As mentioned above, it turns out that in ${}^{87}$Rb, because the singlet 
and triplet scattering  phase shifts are nearly degenerate, the spin-exchange loss rate is on the order of 
$10^{-14} \rm{cm}^3/s$, so multicomponent ${}^{87}$Rb gases can have lifetimes of more than a
second~\cite{Myatt1997a,Julienne1997a,Kokkelmans1997a,Burke1997a}.
By preparing  ${}^{87}$Rb atoms in a superposition of the hyperfine states $(2,1)$ and $(1,-1)$, this two-level
system in many respects resembles spin polarized hydrogen,
in which the two states correspond to the up and down states of the nuclear spin.
Conceptually, this system is a cross between a spinor gas and a binary mixture of distinct atomic species.
Under many circumstances, it can be appropriately treated as a binary mixture because,
in the absence of an external coupling field, there are no {\it{intrinsic}} mechanisms that lead to
interconversion of the two spin states, so that the relative spin population is conserved
\footnote{We emphasize that the relative population is conserved {\it{globally}}, however, not {\it{locally}}.}.
Unlike in a mixture of two distinct atomic species, however, 
the relative phase between hyperfine states can play an important role in the collective
dynamics due to the exchange interaction discussed above.
At $T=0$, the exchange interaction is absent from the coupled Gross-Pitaevskii (GP) equations
describing the two-component condensate, 
and so the system appears equivalent to a binary mixture
\footnote{Another way to say this is that the mean field interaction only depends on the densities
$|\Phi_i({\bf{r}},t)|^2$ and not on the cross terms $\Phi_i^*({\bf{r}},t)\Phi_j({\bf{r}},t)$,
which are absent in the coupled GP equations}.
At finite temperatures, however, due to the exchange mean field, the relative phase between the components can
strongly affect the collective dynamics of the relative density, which makes the system absolutely distinct
from a binary mixture.
Put another way, a spin-1/2 system has an extra degree of freedom---the transverse spin---which is absent from
a binary mixture.

This point may not have been fully appreciated before the recent observation of spin waves at
JILA~\cite{Lewandowski2002a,McGuirk2002a} because all previous 
experiments~\cite{Hall1998b,Matthews1999b,Matthews1999a,Matthews1999c} 
and most theoretical treatments of the two-component ${}^{87}$Rb gas 
had concentrated mainly on zero temperature behavior of the condensate.
In contrast to these earlier works, the recent JILA experiments were done in the high-temperature
regime $T\sim 2T_c$, where the quantum degeneracy has little effect on the thermodynamic
properties of a gas.
All the atoms are initially prepared in the same (1,-1) state
(i.e., all in the spin-up state), and
then a $\pi/2$ pulse is applied to tip the spins into the transverse direction.
The spin vector then precesses about the longitudinal axis at a rate proportional to
the energy difference between hyperfine states. 
Due to the mean field and differential Zeeman effects, the local frequency splitting between
hyperfine states varies approximately quadratically with position.
This inhomogeneity initiates collective spin dynamics through the exchange mean field.
A large inhomogeneity induces nonlinear spin oscillations, which cause the striking
spin-state segregation initially observed \cite{Lewandowski2002a}.
This inhomogeneous frequency splitting can be also made arbitrarily small to study the linear
response of the system.  
This technique was used to probe intrinsic collective spin oscillations \cite{McGuirk2002a}.  
The collective spin dynamics observed at JILA above $T_c$ are well understood by the theory
based on the Boltzmann kinetic equation with spin degree of freedom
\cite{Oktel2002b,Fuchs2002a,Williams2002a,Nikuni2002a}. 

One exciting advancement of the recent spin wave experiments at JILA
\cite{Lewandowski2002a,McGuirk2002a} is the technique
to obtain spatially resolved images of spin dynamics in a gas. 
The density profile of either spin state is measured using absorption imaging.
Together with the Ramsey fringe technique, integrated spatial profiles
of the longitudinal and transverse phase angles of the local spin density
can be extracted from experimental data, as shown in the stunning images in
McGuirk {\it{et al.}} \cite{McGuirk2002a}. 
This is in sharp contrast to the earlier experiments on the spin-polarized gases,
where NMR was done on the system to obtain the frequency spectrum spatially averaged over the 
entire sample. 
This new technique stimulates theoretical investigation of the spatial structure of the spin
dynamics~\cite{Nikuni2002a} that can be directly compared with experimental observations~\cite{McGuirk2002a}.

\subsection{Outline of this paper}
In this paper we derive coupled dynamical equations for the condensate and noncondensate
components, both of which have spin-1/2 internal degrees of freedom.
For a single-component Bose-condensed gas,
finite-temperature kinetic theory has been a major field of
theoretical study, and several groups have derived kinetic theories 
using different approaches
\cite{Stoof1999a,Walser1999a,Zaremba1999a,Gardiner2001a}.
Among them, our present theory has the most overlap with the kinetic theory
by Zaremba, Nikuni, and Griffin (ZNG) \cite{Zaremba1999a},
which uses the semiclassical description of the noncondensate atoms within the Hartree-Fock
approximation---a theory closely related to the pioneering work of Kirkpatrick and
Dorfman \cite{Kirkpatrick1985c}.
Recent numerical simulations by Jackson and Zaremba \cite{Jackson2002a} showed that
the ZNG kinetic theory provides an excellent quantitative description of various
dynamical phenomena in trapped Bose gases at finite temperatures.
It is thus natural to also expect the same kind of semiclassical kinetic theory to be
effective for a spin-1/2 Bose-condensed gas.
In Section \ref{sec:derivation}, we give a detailed derivation of
a generalized GP equation for the condensate and a semiclassical Boltzmann kinetic equation for
noncondensate atoms.
These two components are coupled through the Hartree-Fock mean field as well as collisions
(corresponding to ``$C_{12}$" collisions in the ZNG theory).
In the ZNG theory, $C_{12}$ collisions play an important role in collisional damping of the
condensate collective modes \cite{Williams2001b},
condensate growth \cite{Bijlsma2000a,Davis2000a},
and vortex nucleation \cite{Williams2002b}.
We will show in this paper that the condensate-noncondensate collisions also play an 
important role in the spin-1/2 gas in bringing the two spin components into local equilibrium with
each other.

We emphasize that the dynamics of the spin-1/2 Bose-condensed gas at finite
temperature is potentially more interesting than that of a single-component system because the
transverse spin of the thermal cloud can have strong
collective dynamics even in the very dilute limit. 
In contrast, the mean field of the noncondensate plays a very minor role on the density fluctuations
in a single-component Bose gas, (mainly as a source of damping of the condensate
excitations) \cite{Zaremba1999a,Jackson2002a}. 
To gain more physical insight into the coupled spin dynamics,
in Section \ref{sec:hydro} we recast the kinetic theory derived in Section \ref{sec:derivation}
into a form of spin hydrodynamic equations, which are written in terms of the spin
and spin current for the condensate and noncondensate components.
At finite temperatures, the spins of the condensate and thermal gas interact strongly
through the exchange mean field.
The role of the exchange interaction between the two components has been discussed by
Oktel and Levitov \cite{Oktel2002c} in a spatially homogeneous system.
Our hydrodynamic equations also involve the collisional relaxation term due to ``$C_{12}$-type"
collisions, which try to bring the thermal cloud and condensate spins into local equilibrium.

In Section \ref{sec:different_g}, we investigate the effect of different scattering
lengths on the relaxation of the transverse spin polarization.
A similar discussion was given in a recent paper by Bradley and Gardiner
\cite{Bradley2002a}.
In Section \ref{sec:conclusions}, we summarize this paper,
and discuss some possible future applications of our kinetic theory.
All the detailed calculations of the collision integrals and transport relaxation
times are given in the Appendices.

\section{Derivation of the spin-1/2 kinetic equations}
\label{sec:derivation}

We consider a trapped Bose-condensed gas of atoms with two internal states, which are denoted as
$|1\rangle$ and $|2\rangle$.
For example, these can be any of the systems shown in Fig.~1 as discussed in the Appendix.
The state of a single atom is given by a spinor wavefunction
\begin{equation}
\underline \phi({\bf r},t)=\left(\matrix{ \phi_1({\bf r},t) \cr
\phi_2({\bf r},t) \cr}\right) = \phi_1({\bf r},t)|1\rangle + 
\phi_2({\bf r},t)|2\rangle.
\end{equation}
Bold-faced variables are vectors in coordinate space and a single underline indicates a spinor variable.
The single-atom Hamiltonian is given by
\begin{equation}
\uuline{H}{}_0({\bf r}) = \left[-\frac{\hbar^2}{2m}\nabla^2+U_{\rm
ext}({\bf r})\right]\uuline{1} + \frac{\hbar}{2}\vec{\Omega}({\bf r})
\cdot\vec{\uuline{\sigma}},
\end{equation}
where the double underline indicates a $2\times2$ matrix, and the vector symbol (e.g. $\vec{\Omega}$)
indicates an SU(2) spin vector. The first term
describes the center-of-mass motion of an atom in a harmonic trap
\begin{equation}
U_{\rm ext}({\bf r}) = \frac{1}{2}m 
\left( \omega_x^2x^2 + \omega_y^2y^2 + \omega_z^2z^2 \right).
\end{equation}
The second term describes the evolution of the spin state of the atom.
The dot product can be expanded as 
\begin{equation}
\vec{\Omega}({\bf r}) \cdot\vec{\uuline{\sigma}}
= \Omega^x({\bf r}) \uuline{\sigma}^x + 
  \Omega^y ({\bf r})\uuline{\sigma}^y + 
  \Omega^z({\bf r}) \uuline{\sigma}^z.
\end{equation}
The matrices $\uuline{\sigma}{}^j~(j=x,y,z)$ are the usual Pauli spin matrices.
The quantity $\vec\Omega({\bf r})$ describes an external field (analogous
to an external magnetic field in a true spin-1/2 system). The
transverse components $\Omega^x({\bf r})$ and $\Omega^y({\bf r})$
describe the possible coupling between the two spin states (for
example, if a microwave field or laser is applied that is tuned close
to resonance) and the longitudinal component $\Omega^z({\bf r})$
describes the energy offset between the two internal states. 
In addition to being position dependent, the harmonic trap $U_{\rm{ext}}$ and the external field
$\vec\Omega$ may also vary in time.
We describe the collisions between two atoms in the gas by the
contact interaction term
\begin{eqnarray}
V_{\otimes 2}({\bf r}, {\bf r}') &=& \delta({\bf r}-{\bf r}')
\Big [V_0 \uuline{1}\otimes \uuline{1}
+ V_z (\uuline{1}\otimes \uuline{\sigma}^z + 
\uuline{\sigma}^z\otimes \uuline{1})
+ V_{zz} \uuline{\sigma}^z\otimes\uuline{\sigma}^z \Big ] \cr
&\equiv&\delta({\bf r}-{\bf r}')V_{\otimes 2},
\label{Vtensor}
\end{eqnarray}
where $\otimes$ is the tensor product.
The subscript $\otimes 2$ denotes a $4 \times 4$ matrix formed from the tensor product of spin
matrices and $V_0 = (g_{11} + g_{22} +
2 g_{12})/4$, $V_z = (g_{11}-g_{22})/4$, and $V_{zz} =(g_{11}+g_{22}-2g_{12})/4$. 
The tensor $V_{\otimes 2}$ has the following four components:
\begin{equation}
\langle i |\langle j|V_{\otimes 2}|i\rangle |j\rangle=g_{ij},
~~(i,j=1,2).
\end{equation}
The interaction strengths for the various
collision processes have the form
\begin{equation}
g_{ij} = \frac{4 \pi \hbar^2 a_{ij}}{m},
\end{equation}
where $a_{ij}$ are the $s$-wave scattering lengths describing collisions 
(e.g. $|i\rangle  |j\rangle\rightarrow |i\rangle |j\rangle$).
The scattering lengths for a specific two-level system must be obtained from a detailed multichannel 
scattering calculation. 
For the case of ${}^{87}$Rb prepared in the states $|1\rangle = |F=1,M_F = -1\rangle$ and
$|2\rangle = |F=2,M_F = 1\rangle$, 
these scattering lengths have been determined to be $a_{11}=100.9 a_0$, $a_{12}=98.2 a_0$,
$a_{22}=95.6 a_0$, where $a_0$ is the Bohr radius \cite{Lewandowski2002a}.
The near degeneracy of these scattering lengths gives the result $V_{zz}\approx 0$ in
Eq.~(\ref{Vtensor}) and the first term proportional to $V_0$ dominates
over the second term. If one makes the approximation of equal scattering lengths,
then there is no explicit spin dependence in the interaction $V_z=V_{zz}=0$.

\begin{figure}
  \centerline{\epsfig{file=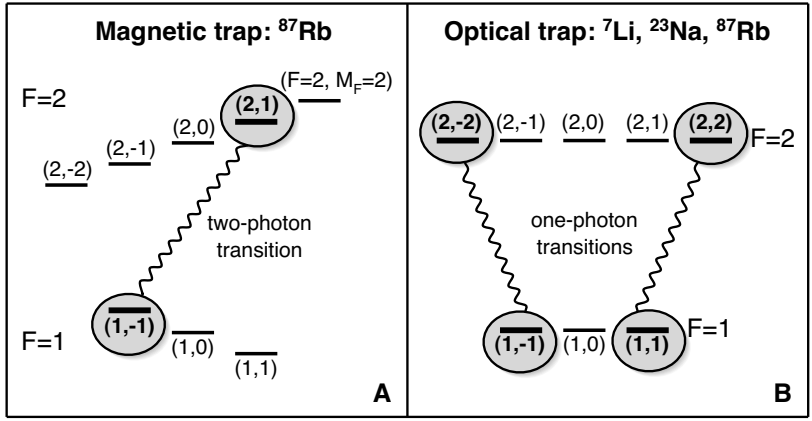,height=2.3in}}
\caption{Several possible two-level systems. Panel A shows the two states used in the 
JILA experiments, where ${}^{87}$Rb atoms are held in a magnetic trap. Panel B shows
two different cases that would make good two-level systems for the alkali atoms confined in an
optical trap.}
\end{figure}

In quantum field theory, the many-body Hamiltonian of this
system is given by
\begin{eqnarray}
\hat{H}=\sum_{ij}\int d{\bf r}\hat\psi^{\dagger}_i({\bf r})
\langle i |\uuline{H}{}_0({\bf r})|j\rangle \hat\psi_j({\bf r})
+\sum_{ij}\frac{g_{ij}}{2}
\int d{\bf r}\hat\psi^{\dagger}_i({\bf r}) \hat\psi^{\dagger}_j({\bf r}) 
\hat\psi_j({\bf r}) \hat\psi_i({\bf r}),
\end{eqnarray}
where $\hat\psi_i({\bf r})$ is a Bose field operator satisfying the
commutation relation
\begin{equation}
[\hat\psi_i({\bf r}), \hat\psi_j^\dagger({\bf r}')]=\delta_{ij}\delta({\bf
r}-{\bf r}').
\end{equation}
The hat indicates a second quantized operator. 
An important property of this Hamiltonian is that, in the absence of the transverse coupling field
(i.e. $\Omega^x=\Omega^y=0$), the total number operator for each state, given by 
\begin{equation} \hat{N}_i  = \int d{\bf r} \hat\psi^{\dagger}_i({\bf r})\hat\psi_i({\bf r}),
\end{equation}
commutes with $\hat{H}$, which means that the population of each state is separately conserved.
This property, which also occurs in a binary mixture of distinct atomic species,
means that the two levels will not become thermally populated in equilibrium 
(in our model, there are no state-changing collisions that lead to a net interconversion between
states).

We next define the time evolution of the system in terms of the
statistical density operator, and introduce the Heisenberg
representation for the field operators.  The dynamics of the system is
described by the density operator $\hat\rho(t)$, from which one can
obtain the expectation value of an arbitrary operator $\hat O$ (which has
no explicit time dependence)
\begin{equation}
\langle\hat O\rangle_t={\rm tr}\hat\rho(t)\hat O.
\label{Ot}
\end{equation}
The state of the many body system evolves in time according to
\begin{equation}
i\hbar\frac{d}{dt}\hat\rho(t)=[\hat H,\hat\rho(t)].
\label{Liouville}
\end{equation}
With Eq.~(\ref{Ot}) and Eq.~(\ref{Liouville}), the equation of motion for
the quantity $\langle\hat O\rangle_t$ is given by
\begin{equation}
i\hbar\frac{d}{dt} \langle \hat O \rangle_t=
\langle [\hat O,\hat H]\rangle_t.
\end{equation}

It is convenient to introduce the time evolution operator 
$\hat\EuScript{U}(t,t_0)$, 
which obeys the equation of motion
\begin{equation}
i\hbar\frac{d}{dt}\hat \EuScript{U}(t,t_0)=\hat H\hat \EuScript{U}(t,t_0),
\label{hatScriptU}
\end{equation}
with $\hat \EuScript{U}(t_0,t_0)=1$. 
Here $t_0$ is the time at which the initial nonequilibrium density
matrix $\hat\rho(t_0)$ is specified.
One can then express the time evolution of $\hat\rho(t)$ as
\begin{equation}
\hat\rho(t)=\hat \EuScript{U}(t,t_0)\hat\rho(t_0)
\hat \EuScript{U}^{\dagger}(t,t_0).
\end{equation}
Thus, the time evolution of $\langle\hat O\rangle_t$ can be written as
\begin{equation}
\langle \hat O \rangle_t=
{\rm tr}\hat \EuScript{U}(t,t_0)\hat\rho(t_0)
\hat \EuScript{U}^{\dagger}(t,t_0)\hat O
={\rm tr}\hat\rho(t_0)\hat \EuScript{U}^{\dagger}(t,t_0)\hat O
\hat \EuScript{U}(t,t_0)
\equiv\langle\hat O(t)\rangle,
\end{equation}
where $\hat O(t)\equiv \hat \EuScript{U}^{\dagger}(t,t_0)\hat O \hat
\EuScript{U}(t,t_0)$ is the operator in the Heisenberg picture,
which obeys the Heisenberg equation of motion
\begin{equation}
i\hbar\frac{\partial \hat O(t)}{\partial t}
=[\hat O(t),\hat H(t)].
\end{equation}

The Heisenberg equation of motion for the Bose field operator is then
given by
\begin{eqnarray}
i\hbar\frac{\partial}{\partial t}\hat\psi_i({\bf r},t)
&=&\left[-\frac{\hbar^2\nabla^2}{2m}+ U_{\rm ext}({\bf
r})\right]\hat\psi_i({\bf r},t) +\sum_j
g_{ij}\hat\psi^{\dagger}_j({\bf r},t)\hat\psi_j({\bf r},t)
\hat\psi_i({\bf r},t) \cr &&+\sum_j\frac{\hbar}{2}\langle i
|\vec{\Omega}({\bf r},t)\cdot\vec{\uuline{\sigma}}|j\rangle
\hat\psi_j({\bf r},t).
\label{eq_psi}
\end{eqnarray}

The key idea in capturing the physics the Bose-Einstein condensation
is to introduce the broken-symmetry order parameter
\begin{equation}
\underline \Phi({\bf r},t)=\left(\matrix{ \Phi_1({\bf r},t) \cr
\Phi_2({\bf r},t) \cr}\right) = \Phi_1({\bf r},t)|1\rangle + 
\Phi_2({\bf r},t)|2\rangle,
\label{Phi_spinor}
\end{equation}
with
\begin{equation}
\Phi_i({\bf r},t)=\langle\hat\psi_i({\bf r})\rangle_t=
\langle\hat\psi_i({\bf r},t)\rangle.
\label{Phi_i}
\end{equation}
In addition to the usual U(1) gauge symmetry breaking, the order
parameter defined by Eq.~(\ref{Phi_spinor}) also breaks the SU(2) symmetry,
which is associated with the spin-1/2 internal degree of freedom.  
The noncondensate field operator is defined by
\begin{equation}
\tilde\psi_i({\bf r},t)\equiv\hat\psi_i({\bf r},t)-\Phi_i({\bf r},t),
\end{equation}
with $\langle\tilde\psi({\bf r},t)\rangle=0$.  
This noncondensate operator satisfies the equal-time commutation relation
\begin{equation}
[\tilde\psi_i({\bf r},t),\tilde\psi_j^{\dagger}({\bf r}',t)]=\delta_{ij}\delta({\bf r}-{\bf r}').
\end{equation}
The equation of motion for $\Phi_i$ can be obtained by taking the expectation value of
Eq.~(\ref{eq_psi}), which yields
\begin{eqnarray}
i\hbar\frac{\partial\Phi_i({\bf r},t)}{\partial t}
&=&\sum_j \Biggl ( \left \{ \left[ -\frac{\hbar^2\nabla^2}{2m}+
U_c({\bf r},t)\right] \delta_{ij}
+\frac{\hbar}{2}\langle i |\vec{\Omega}_c\cdot
\vec{\uuline{\sigma}}|j\rangle\right \{
\Phi_j({\bf r},t) \cr
&&+g_{ij}\left[\tilde m_{ij}({\bf r},t)\Phi_j^*({\bf r},t)
+\langle \tilde\psi^{\dagger}_j({\bf r},t)\tilde\psi_j({\bf r},t)
\tilde\psi_i({\bf r},t)\rangle\right]\Biggr ).
\label{d_Psi_i_dt}
\end{eqnarray}
Here we define the effective potential $U_c$ and coupling field
$\vec{\Omega}_c$, which include the mean-field interaction of
the condensate with itself and with the noncondensate
\begin{equation}
U_c=U_{\rm ext}+\frac{1}{2}
[g_{11}n_1+g_{22}n_2+g_{12}(n_1+n_2)+
g_{11}\tilde n_1+g_{22}\tilde n_2],
\end{equation}
\begin{equation}
\vec{\Omega}_c=\vec{\Omega}'_c+\frac{g_{12}}{\hbar}{\rm Tr}
\vec{\uuline{\sigma}}\,\uuline{\tilde n},
\label{Omegac} 
\end{equation}
where ``${\rm Tr}$'' denotes the spin trace.
The condensate and noncondensate densities are defined by
\begin{equation}
\langle i |\uuline{n}{}_c({\bf r},t) | j \rangle = 
n_{cij}({\bf r},t)\equiv \Phi_j^*({\bf r},t)\Phi_i({\bf r},t),
\label{ncdef}
\end{equation}
\begin{equation}
\langle i |\tilde \uuline{n}{}({\bf r},t) | j \rangle = 
\tilde n_{ij}({\bf r},t)\equiv\langle\tilde\psi^{\dagger}_j({\bf r},t)
\tilde\psi_i({\bf r},t)\rangle,
\label{nndef}
\end{equation}
with $n_{ci}\equiv n_{cii}$ and $\tilde n_{i}\equiv\tilde n_{ii}$.
The total density of internal state $i$ is $n_i \equiv n_{ci} + \tilde n_{i}$.
The second term in Eq.~(\ref{Omegac}) directly couples the spins of the condensate and
noncondensate and is present even when the interaction is independent of spin (i.e. when the 
scattering lengths are equal).
It is similar to the molecular field
of a magnetically ordered system and plays a
central role in the collective spin dynamics.
At $T=0$, however, when the noncondensate is absent, this term vanishes.
The effect of different scattering lengths is to modify the longitudinal component of the external 
field 
\begin{equation}
\vec{\Omega}'_c =(\Omega^x,\Omega^y,\Omega^z+ 2\Delta_n+\Delta_c).
\label{Omegacp}
\end{equation}
The mean field frequency shifts (or ``clock" shifts) due to the condensate $\Delta_c$ and
noncondensate $\Delta_n$ (which is multiplied by the Bose-enhanced factor of $2$ in 
Eq.~(\ref{Omegacp})~\cite{Busch2000a}) are
\begin{equation}
\Delta_c=[g_{11}n_{c1}+g_{12}n_{c2}-(g_{22}n_{c2}+g_{12}n_{c1})]
/\hbar.
\label{Delta_c}
\end{equation}
\begin{equation}
\Delta_n=[g_{11}\tilde n_1+g_{12}\tilde n_2-(g_{22}\tilde n_2+g_{12}\tilde n_1)]
/\hbar,
\label{Delta_n}
\end{equation}
Finally, the anomalous correlation appearing in Eq.~(\ref{d_Psi_i_dt}) of the noncondensate
is defined by
\begin{equation}
\tilde m_{ij}({\bf r},t)\equiv\langle\tilde\psi_j({\bf r},t)
\tilde\psi_i({\bf r},t)\rangle.
\end{equation}
The three field operator average 
$\langle \tilde\psi^{\dagger}_j({\bf r},t)\tilde\psi_j({\bf r},t)\tilde\psi_i({\bf r},t)\rangle$ 
in Eq.~(\ref{d_Psi_i_dt}) includes higher order correlations that will give rise to a binary
collision integral describing the exchange of atoms between the condensate and the noncondensate.
Before evaluating it explicitly, we next consider the dynamics of the
noncondensate.

The equation of motion for the noncondensate field operator
$\tilde\psi_i$ can be derived directly from Eq.~(\ref{eq_psi})
and Eq.~(\ref{d_Psi_i_dt}): 
\begin{eqnarray}
i\hbar\frac{\partial\tilde\psi_i}{\partial t}
&=&\left(-\frac{\hbar^2\nabla^2}{2m}+U_n\right)\tilde\psi_i
+\sum_j\frac{\hbar}{2}\langle i |\vec{\Omega}_n\cdot
\vec{\uuline{\sigma}}|j\rangle \tilde\psi_j \cr
&-&\sum_jg_{ij}\left(\Phi_i\tilde n_{jj}+\Phi_j\tilde n_{ij}+\Phi^{*}_j\tilde m_{ij}
+\langle \tilde\psi_j^{\dagger}\tilde\psi_j\tilde\psi_i\rangle\right) \cr
&+&\sum_{j}g_{ij}
(\Phi_i\tilde\psi^{\dagger}_j\tilde\psi_j+\Phi_j\tilde\psi^{\dagger}_j\tilde\psi_i
+\Phi^{*}_j\tilde\psi_j\tilde\psi_i) \cr
&+&\sum_{j}g_{ij}(\tilde\psi^{\dagger}_j\tilde\psi_j\tilde\psi_i
-\tilde n_{jj}\tilde\psi_i-\tilde n_{ij}\tilde\psi_j),
\label{tilde_psi_dt}
\end{eqnarray}
where the effective potential $U_n({\bf r},t)$ and coupling field
$\vec{\Omega}_n({\bf r},t)$ (analogous to $U_c$ and $\vec{\Omega}_c$ for the condensate) include
the mean-field interaction of the noncondensate with itself and with the condensate:
\begin{equation}
U_n\equiv U_{\rm ext}+g_{11}n_1+g_{22}n_2+\frac{g_{12}}{2}(n_1+n_2),
\label{def_un}
\end{equation}
\begin{equation}
\vec{\Omega}_n=\vec{\Omega}'_n+\frac{g_{12}}{\hbar}
{\rm Tr} \uuline{\vec{\sigma}}(\uuline{n}{}_c+\uuline{\tilde n}).
\label{Omegan}
\end{equation}
The modified external field including the mean-field frequency shifts (or clock shifts) due to different
scattering lengths is given by
\begin{equation}
\vec{\Omega}'_n = (\Omega^x,\Omega^y,\Omega^z +2\Delta_n+2\Delta_c),
\label{Omegapn}
\end{equation}
where $\Delta_c$ and $\Delta_n$ were defined in Eq.~(\ref{Delta_c}) and Eq.~(\ref{Delta_n}).
We note that this modified field for the noncondensate differs from Eq.~(\ref{Omegacp}) for the
condensate by the factor of 2 in front of $\Delta_c$. These clock shifts were measured in recent
experiments \cite{Harber2002a}.
We also note that the second term in Eq.~(\ref{Omegan}) differs from that found for the condensate. 
In Eq.~(\ref{Omegac}) for the condensate, the noncondensate spin couples to the condensate due
to the exchange term in the mean field interaction,
but the condensate itself does not contribute to this effect.
In contrast, in Eq.~(\ref{Omegan}) both the condensate and noncondensate contribute to this term.
Above $T_c$, when the condensate is absent, this spin mean field term is still present and plays a
dominant role in the collective spin dynamics of the noncondensate 
(it is in fact responsible for transverse spin waves in the thermal cloud 
\cite{Williams2002a,Nikuni2002a}). 

It is useful to introduce a time evolution operator
$\tilde\EuScript{U}(t,t_0)$ (distinct from $\hat\EuScript{U}(t,t_0)$
given in Eq.~(\ref{hatScriptU})) that evolves $\tilde\psi_i$ in time
according to Eq~(\ref{tilde_psi_dt})
\begin{equation}
\tilde\psi_i({\bf r},t)=\tilde 
\EuScript{U}^{\dagger}(t,t_0)\tilde\psi_i({\bf r},t_0)
\tilde \EuScript{U}(t,t_0).
\label{U_tildepsi_U}
\end{equation}
This operator $\tilde \EuScript{U}(t,t_0)$ evolves according to the
equation of motion
\begin{equation}
i\hbar\frac{d}{dt}\tilde \EuScript{U}(t,t_0)=\hat H_{\rm eff}
\tilde \EuScript{U}(t,t_0),
\label{tildeU}
\end{equation}
where the effective Hamiltonian $\hat H_{\rm eff}$ is defined as
\begin{equation}
\hat H_{\rm eff}=\hat H_0+\hat H',
\label{Heff}
\end{equation}
\begin{equation}
\hat H'=\hat H'_1+\hat H'_2+\hat H'_3+\hat H'_4.
\label{Hprime}
\end{equation}
In writing Eq~(\ref{Heff}), we have separated the Hamiltonian into a
part that describes the mean-field dynamics $\hat H_0$ and the
remaining part $\hat H'$, which can be treated perturbatively.
The mean field part is given by
\begin{equation}
\hat H_0=\sum_{ij}\int d{\bf r}\tilde\psi^{\dagger}_i
\Big\langle i\Big|\left (-\frac{\hbar^2}{2m}\nabla^2+U_n\right )\uuline{1}
+\frac{\hbar}{2}
\vec{\Omega}_n\cdot\vec{\uuline{\sigma}} \Big|j\Big\rangle\tilde\psi_j, 
\end{equation}
where the effective potential $U_n$ and coupling field $\Omega_n$ have been
defined in Eqs.~(\ref{def_un})-(\ref{Omegapn}).

The four terms contributing to $H'$ are
\begin{eqnarray}
\hat H_1'&=&-\sum_{ij}g_{ij}\int d{\bf r}\left[\left(\Phi_i\tilde n_{jj}
+\Phi_j\tilde n_{ij}+\Phi^{*}_j\tilde m_{ij}
+\langle\tilde\psi_j^{\dagger}\tilde\psi_j\tilde\psi_i\rangle \right)
\tilde\psi_i^{\dagger}+{\rm H.c.}\right],\\
\hat H_2'&=&\sum_{ij}\frac{g_{ij}}{2}\int d{\bf r}(\Phi^{*}_i\Phi^{*}_j
\tilde\psi_j\tilde\psi_i
+\Phi_i\Phi_j\tilde\psi^{\dagger}_j\tilde\psi^{\dagger}_i),\\
\hat H_3'&=&\sum_{ij}g_{ij}\int d{\bf r}(\Phi^*_i\tilde\psi^{\dagger}_j
\tilde\psi_j\tilde\psi_i+\Phi_i\tilde\psi^{\dagger}_i\tilde\psi^{\dagger}_j
\tilde\psi_j), \\
\hat H_4'&=&\sum_{ij}\int d{\bf r}\left[
\frac{g_{ij}}{2}\tilde\psi^{\dagger}_i\tilde\psi^{\dagger}_j\tilde\psi_j
\tilde\psi_i-g_{ij}(\tilde n_{ji}\tilde\psi^{\dagger}_j\tilde\psi_i
+\tilde n_{jj}\tilde\psi^{\dagger}_i\tilde\psi_i)\right].
\label{H4}
\end{eqnarray}
In the above formulas, the arguments of the field operators are $({\bf r},t_0)$, i.e.,
$\tilde\psi\equiv\tilde\psi({\bf r},t_0)$. 
Note that the effective Hamiltonian $\hat H_{\rm eff}$ has time dependence
through $\tilde n_{ij}({\bf r},t),\Phi_i({\bf r},t)$, and $\tilde m_{ij}({\bf
r},t)$ (apart from the possible time variation of the external
potentials), which we have suppressed so as not to confuse these
quantities with the Heisenberg operators. We emphasize here that we
have made no additional approximations in introducing $\tilde\EuScript{U}$, 
that is, one can show that
Eqs.~(\ref{Heff})-(\ref{H4}), together with Eqs.~(\ref{U_tildepsi_U})
and (\ref{tildeU}), reproduce the original Heisenberg equation of motion
in Eq.~(\ref{tilde_psi_dt}).

For an operator $\hat O(t)$ constructed from the noncondensate field
operators $\tilde\psi_i({\bf r},t)$ and
$\tilde\psi^{\dagger}_i({\bf r},t)$, one can show that
\begin{eqnarray}
{\rm tr}\hat\rho(t_0)\hat O(t)&=&{\rm tr}\hat\rho(t_0)
\tilde \EuScript{U}^{\dagger}(t,t_0)\hat O(t_0) \tilde \EuScript{U}(t,t_0) \cr
&=&{\rm tr}\tilde \EuScript{U}(t,t_0)\hat\rho(t_0)\tilde \EuScript{U}^{\dagger}(t,t_0)
\hat O(t_0) \cr
&\equiv& {\rm tr}\tilde \rho(t,t_0)\hat O(t_0) \cr
&\equiv& \langle \hat O \rangle_t.
\label{Oavgt}
\end{eqnarray}
This defines the effective {\it{noncondensate}} density operator
\begin{equation}
\tilde\rho(t,t_0)\equiv\tilde \EuScript{U}(t,t_0)\hat\rho(t_0)\tilde \EuScript{U}^{\dagger}(t,t_0),
\end{equation}
which obeys the equation of motion
\begin{equation}
i\hbar\frac{d}{dt}\tilde\rho(t,t_0)=[\hat H_{\rm eff}(t),\tilde\rho(t,t_0)].
\label{eq_rhotilde}
\end{equation}

Having defined the time evolution of the noncondensate field operator,
we now turn to the derivation of the kinetic equation for the
noncondensate. We introduce the Wigner operator
\begin{equation}
\hat W_{ij}({\bf r},{\bf p},t)\equiv \int d{\bf r}'
e^{i{\bf p}\cdot{\bf r}'/\hbar}\tilde\psi^{\dagger}_j({\bf r}+{\bf r}'/2,t)
\tilde\psi_i({\bf r}-{\bf r}'/2,t),
\end{equation}
and define the semiclassical distribution function
\begin{equation}
W_{ij}({\bf r},{\bf p},t)\equiv 
\langle i | \uuline{W}({\bf r},{\bf p},t) | j \rangle
\equiv{\rm tr}\hat \rho(t_0)
\hat W_{ij}({\bf r},{\bf p},t)
={\rm tr}\tilde \rho(t,t_0)\hat W_{ij}({\bf r},{\bf p},t_0).
\end{equation}
Knowledge of this function allows one to calculate various
nonequilibrium expectation values, such as the noncondensate density
introduced in Eq.~(\ref{nndef})
\begin{equation}
\tilde n_{ij} ({\bf r},t)= \int \frac{d{\bf p} }{ (2\pi\hbar)^3}
W_{ij}({\bf r},{\bf p},t).
\label{tilde_n}
\end{equation}
The equation of motion for $W_{ij}$ given from Eq.~(\ref{eq_rhotilde}) is
\begin{eqnarray}
\frac{\partial}{\partial t}W_{ij}({\bf r},{\bf p},t)
&=&
\frac{1}{i\hbar}{\rm tr}\tilde\rho(t,t_0)
[\hat W_{ij}({\bf r},{\bf p},t_0),\hat H_{\rm eff}(t)] \cr
&=&
\frac{1}{i\hbar}{\rm tr}\tilde\rho(t,t_0)
[\hat W_{ij}({\bf r},{\bf p},t_0),\hat H_0(t)]  \cr
&&+ \frac{1}{i\hbar}{\rm tr}\tilde\rho(t,t_0)
[\hat W_{ij}({\bf r},{\bf p},t_0),\hat H'(t)].
\end{eqnarray}
The first term on the right hand side defines the free-streaming
operator in the kinetic equation. With the assumption that $U_n({\bf
r},t)$ and $\vec{\Omega}_n({\bf r},t)$ vary slowly in space, then we
can neglect the quantum corrections to the free-streaming term~\cite{Zurek1994a},
which leads to the semi-classical result
\begin{equation}
\frac{\partial\uuline{W} } {\partial t}+
\frac{\bf p}{m}\cdot{\boldsymbol\nabla}_r\uuline{W}
-\frac{1}{2}\{{\boldsymbol\nabla}_r\uuline{U},{\boldsymbol\nabla}_p\uuline{W}\}
-\frac{i}{\hbar}[\uuline{W},\uuline{U}]=\uuline{I},
\label{kine_eq1}
\end{equation}
where $[~,~]$ and $\{~,~\}$ on the left-hand side represent the commutator and
anticommutator for the $2 \times 2$ matrices. 
Here we have defined the $2 \times 2$ matrix for the noncondensate effective potential
\begin{equation}
\uuline{U}({\bf r},t)\equiv U_n({\bf r},t)\uuline{1}
+\frac{\hbar}{2}\vec{\Omega}_n({\bf r},t)\cdot\vec{\uuline{\sigma}}.
\end{equation}
The term on the right-hand side is given by 
\begin{equation}
\langle i | \uuline{I} | j
\rangle\equiv\frac{1}{i\hbar}{\rm tr}\tilde\rho(t,t_0) [\hat W_{ij}({\bf
r},{\bf p},t_0),\hat H'(t)].
\end{equation}

To summarize the derivation so far, we now have the condensate
equation of motion Eq.~(\ref{d_Psi_i_dt}) and the kinetic equation for
the noncondensate Eq.~(\ref{kine_eq1}). Both of these equations depend
on higher order correlations, which enter through the three field
average $\langle \tilde\psi^{\dagger}_j({\bf r},t)\tilde\psi_j({\bf
r},t) \tilde\psi_i({\bf r},t)\rangle$ in Eq.~(\ref{d_Psi_i_dt}) and
through the collision term $\uuline{I}$ in Eq.~(\ref{kine_eq1}).  In order
to obtain a closed set of equations for $\Phi_i({\bf r},t)$ and
$\uuline{W}({\bf r},{\bf p},t)$, we must truncate the description by
making a Markovian ansatz about collisions.  This can be carried out
using the machinery of perturbation theory.

To carry out perturbation theory, we first write down the formal
solution to the equation of motion Eq~(\ref{eq_rhotilde}) for
$\tilde\rho(t,t_0)$
\begin{equation}
\tilde\rho(t,t_0) = \hat \EuScript{U}_0(t,t_0)
\hat \rho(t_0)\hat \EuScript{U}_0^\dagger(t,t_0) 
-\frac{i}{\hbar}\int_{t_0}^{t}dt'\hat \EuScript{U}_0(t,t')
[\hat H'(t'),\tilde \rho(t',t_0)]\hat \EuScript{U}_0^\dagger(t,t'),
\label{formal_sol}
\end{equation}
where $\hat \EuScript{U}_0(t,t_0)$ is the unperturbed evolution operator
\begin{equation}
\hat \EuScript{U}_0(t,t_0)=T\exp\left[-\frac{i}{\hbar}\int_{t_0}^tdt'\hat
H_0(t')\right],
\end{equation}
with $T$ being the time-ordering operator. Iterating
Eq.~(\ref{formal_sol}) to first order in $H'$, one has
\begin{eqnarray}
\tilde\rho(t,t_0) &\simeq& \hat \EuScript{U}_0(t,t_0) \hat \rho(t_0)\hat
\EuScript{U}_0^\dagger(t,t_0) \cr
&-& \frac{i}{\hbar}\int_{t_0}^{t}dt'\hat \EuScript{U}_0(t,t') [\hat
H'(t'),\hat\EuScript{U}_0(t',t_0)\hat
\rho(t_0)\EuScript{U}_0^\dagger(t',t_0)]\hat
\EuScript{U}_0^\dagger(t,t'),
\label{approx_sol}
\end{eqnarray}
Using this in Eq.~(\ref{Oavgt}), a statistical average of an operator $\hat O$
constructed from $\tilde \psi_i$ can be expressed as
\begin{eqnarray}
&&\langle \hat O\rangle_t={\rm tr}\hat\rho(t_0)\Bigl\{
\hat \EuScript{U}_0^{\dagger}(t,t_0)
\hat O(t_0)\hat \EuScript{U}_0(t,t_0) \cr
~~&&-\frac{i}{\hbar}\int_{t_0}^{t}dt'\hat \EuScript{U}_0^{\dagger}(t',t_0)
[\hat \EuScript{U}_0^{\dagger}(t,t')\hat O(t_0)\hat \EuScript{U}_0(t,t'),
\hat H'(t')]\hat \EuScript{U}_0(t',t_0)\Bigr\},
\label{2nd-order}
\end{eqnarray}

We may now use this perturbative solution to calculate the quantities
$\langle \tilde\psi^{\dagger}_j({\bf r},t)\tilde\psi_j({\bf r},t)
\tilde\psi_i({\bf r},t)\rangle$ and $\uuline{I}$, which involve
averages of products of three and four field operators. As in ZNG~\cite{Zaremba1999a},
we effectively calculate these collision integrals to
second order in $g_{ij}$, while explicitly keeping interaction effects
in excitation energy and chemical potentials 
only to first order in $g_{ij}$.
Within this approximation, one can consistently neglect the
anomalous correlation function, i.e.,
we set $\tilde m_{ij}=0$,
and thus there is no contribution from $\hat H_2'(t)$.
A detailed discussion on the anomalous correlation function in a single-component
BEC is found in ZNG \cite{Zaremba1999a}.

In Appendix \ref{sec:collisions}, we provide a detailed derivation showing how the three and four field
operator averages are reduced to binary collision integrals using the
perturbative solution in Eq.~(\ref{2nd-order}). There, we carry out
this calculation for the general case of unequal scattering lengths.
However, for simplicity, and motivated by the near degeneracy of the
scattering lengths in ${}^{87}$Rb, here we focus on the case
$a_{11}=a_{22}=a_{12}\equiv a$ (with $g\equiv 4\pi\hbar^2 a / m$). 
In Section \ref{sec:different_g} we consider how having different scattering lengths modifies the
kinetic equations.

The dynamics of the condensate and noncondensate are described by
equations of motion that are coupled through mean field interaction
terms and collision integrals that describe the exchange of atoms
between the condensate and noncondensate.  The condensate spinor
$\underline\Phi({\bf r},t)$ is described by a generalized finite
temperature Gross-Pitaevskii equation
\begin{eqnarray}
i\hbar\frac{\partial{\underline\Phi}}{\partial t}
&=&\left[\left(-\frac{\hbar^2}{2m}\nabla^2+U_c\right)\uuline{1}
+\frac{\hbar}{2}\vec{\Omega}_c\cdot\vec{\uuline{\sigma}}
-i\uuline{R}\right]{\underline\Phi} \cr
&\equiv&(\uuline{H}{}_c-i\uuline{R})\underline{\Phi}
\label{spinor_GP}
\end{eqnarray}
and the noncondensate atoms are described by the kinetic equation
\begin{eqnarray}
&&\frac{\partial\uuline{W} } {\partial t}
+ \frac{1}{2}  \{\boldsymbol\nabla_p\uuline{H}{}_n,{\boldsymbol\nabla}_r\uuline{W} \}
-\frac{1}{2}\{{\boldsymbol\nabla}_r\uuline{H}{}_n,{\boldsymbol\nabla}_p\uuline{W}\}
-\frac{i}{\hbar}[\uuline{W},\uuline{H}{}_n]\cr
&&~~~~~~~~~~~~~~~~~~~~~~~~~=\uuline{I}{}_c
[\uuline{W}]+\uuline{I}{}_n[\uuline{W}].
\label{KE}
\end{eqnarray}
Here we have defined the condensate Hamiltonian $\uuline{H}{}_c$ and the
semiclassical noncondensate Hamiltonian $\uuline{H}{}_n$ as
\begin{equation}
\uuline{H}{}_c({\bf r},t)\equiv \left[-\frac{\hbar\nabla^2}{2m}+U_c({\bf r},t)\right]
\uuline{1}
+\frac{\hbar}{2}\vec{\Omega}_c({\bf r},t)\cdot\vec{\uuline{\sigma}},
\end{equation}
\begin{equation}
\uuline{H}{}_n({\bf r},{\bf p},t)\equiv
\left[\frac{p^2}{2m}+U_n({\bf r},t)\right]\uuline{1}
+\frac{\hbar}{2}\vec{\Omega}_n({\bf r},t)\cdot\vec{\uuline{\sigma}}.
\end{equation}

The dissipative term $i\uuline{R}$ in the GP equation Eq.~(\ref{spinor_GP}) and
the two collision integrals in the kinetic equation Eq.~(\ref{KE}), 
for the case of equal scattering lengths, can be obtained from the more general expressions
Eqs.~(\ref{In_general2}-\ref{R_general2}) in Appendix \ref{sec:collisions}
\begin{eqnarray}
\uuline{R}&=&g^2\pi \int\frac{d{\bf
p}_1}{(2\pi\hbar)^3} \int \frac{d{\bf p}_2}{(2\pi\hbar)^3}
\int d{\bf p}_3
\delta(\varepsilon_c+\tilde\varepsilon_{p_1}
-\tilde\varepsilon_{p_2}-\tilde\varepsilon_{p_3}) \cr
&&\times\delta({\bf p}_c+{\bf p}_1-{\bf p}_2-{\bf p}_3) \cr
&&\times\bigl\{(\uuline{1}+\uuline{W}{}_3) 
{\rm Tr}[\uuline{W}{}_1 (\uuline{1}+\uuline{W}{}_2)]
-\uuline{W}{}_3{\rm Tr}[(\uuline{1}+\uuline{W}{}_1)\uuline{W}{}_2]
\cr
&&+(\uuline{1}+\uuline{W}{}_2)\uuline{W}{}_1(\uuline{1}+\uuline{W}{}_3)
-\uuline{W}{}_2(\uuline{1}+\uuline{W}{}_1)\uuline{W}{}_3\Bigr\},
\label{R_symmetric}
\end{eqnarray}
\begin{eqnarray}
\uuline{I}{}_n[\uuline{W}]&=&
\frac{g^2\pi}{\hbar}\int\frac{d{\bf p}_2}{(2\pi\hbar)^3}
\int\frac{d{\bf p}_3}{(2\pi\hbar)^3}\int d{\bf p}_4
\delta(\tilde\varepsilon_p+\tilde\varepsilon_{p_2}
-\tilde\varepsilon_{p_3}-\tilde\varepsilon_{p_4}) \cr
&&\times\delta({\bf p}+{\bf p}_2-{\bf p}_3-{\bf p}_4) \cr
&&\times
\Bigl(
\{\uuline{1}+\uuline{W},\uuline{W}{}_4\}
{\rm Tr}[(\uuline{1}+\uuline{W}{}_2)\uuline{W}{}_3] \cr
&&-\{\uuline{W},\uuline{1}+\uuline{W}{}_4\}
{\rm Tr}[\uuline{W}{}_2(\uuline{1}+\uuline{W}{}_3)] \cr
&&+\{\uuline{1}+\uuline{W},\uuline{W}{}_3(\uuline{1}+
\uuline{W}{}_2)\uuline{W}{}_4\}- \cr
&&\{\uuline{W},(\uuline{1}+\uuline{W}{}_3)\uuline{W}{}_2
(\uuline{1}+\uuline{W}{}_4)\}\Bigr),
\label{In_symmetric}
\end{eqnarray}
\begin{eqnarray}
\uuline{I}{}_c[\uuline{W}]&=& \frac{g^2\pi}{\hbar}\int\frac{d{\bf
p}_1}{(2\pi\hbar)^3} \int d{\bf p}_2\int d{\bf p}_3
\delta(\varepsilon_c+\tilde\varepsilon_{p_1}
-\tilde\varepsilon_{p_2}-\tilde\varepsilon_{p_3}) \cr
&&\times\delta({\bf p}_c+{\bf p}_1-{\bf p}_2-{\bf p}_3)
\Bigl\{[\delta({\bf p}-{\bf p}_1)-\delta({\bf p}-{\bf p}_2)] \cr
&&\times\bigl[\{\uuline{1}+\uuline{W}{}_1,\uuline{W}{}_2\} {\rm
Tr}(\uuline{W}{}_3\uuline{n}{}_c)
-\{\uuline{W}{}_1,\uuline{1}+\uuline{W}{}_2\} {\rm
Tr}[(\uuline{1}+\uuline{W}{}_3)\uuline{n}{}_c\bigr] \cr
&&+[(\uuline{1}+\uuline{W}{}_1)\uuline{W}{}_3\uuline{n}{}_c\uuline{W}{}_2
+\uuline{W}{}_2\uuline{n}{}_c\uuline{W}{}_3(\uuline{1}+\uuline{W}{}_1)\cr
&&-\uuline{W}{}_1(\uuline{1}+\uuline{W}{}_3)\uuline{n}{}_c
(\uuline{1}+\uuline{W}{}_2) -(\uuline{1}+\uuline{W}{}_2)\uuline{n}{}_c
(\uuline{1}+\uuline{W}{}_3)\uuline{W}{}_1] \cr &&
-\delta({\bf p}-{\bf p}_3) \bigl(\{\uuline{n}{}_c,\uuline{W}{}_3\}{\rm
Tr}[(\uuline{1}+\uuline{W}{}_1)\uuline{W}{}_2] \cr
&&-\{\uuline{n}{}_c,\uuline{1}+\uuline{W}{}_3\}{\rm
Tr}[\uuline{W}{}_1(\uuline{1}+\uuline{W}{}_2)]\cr
&&+\uuline{n}{}_c\uuline{W}{}_2(\uuline{1}+\uuline{W}{}_1)\uuline{W}{}_3
+\uuline{W}{}_3(\uuline{1}+\uuline{W}{}_1)\uuline{W}{}_2\uuline{n}{}_c\cr
&&-\uuline{n}{}_c(\uuline{1}+\uuline{W}{}_2)\uuline{W}{}_1
(\uuline{1}+\uuline{W}{}_3) -(\uuline{1}+\uuline{W}{}_3)\uuline{W}{}_1
(\uuline{1}+\uuline{W}{}_2)\uuline{n}{}_c]\bigr)\Bigr\},
\label{Ic_symmetric}
\end{eqnarray}
where $\uuline{W}\equiv\uuline{W}({\bf r},{\bf p},t)$ and
$\uuline{W}{}_i\equiv\uuline{W}({\bf r},{\bf p}_i,t)$.

The local condensate and noncondensate energies are defined by
\begin{equation}
\varepsilon_c({\bf r},t)\equiv{\rm Re}[\underline{\Phi}^{\dagger}
\uuline{H}{}_c \underline{\Phi}]/n_c,
\label{ec_def}
\end{equation}
\begin{equation}
\tilde\varepsilon_p({\bf r},t)\equiv {\rm Tr}(\uuline{H}{}_n\uuline{\tilde n})/\tilde n,
\end{equation}
while the condensate momentum is defined by
\begin{equation}
{\bf p}_c\equiv
m{\bf v}_c\equiv \frac{\hbar}{2in_c}\sum_i[\Phi_i^*{\boldsymbol\nabla}\Phi_i-\Phi_i{\boldsymbol\nabla}\Phi_i^*].
\label{vc_def}
\end{equation}
More explicit expressions of the above three quantities are given in 
Eqs.~(\ref{ec_def2}), (\ref{HF_excitation}) and (\ref{vc_def2}) below.

At this stage we pause to discuss various cases of the spin-1/2
kinetic equations (\ref{spinor_GP}) and (\ref{KE}) and make contact
with some earlier work. We first remark that in the case of
$T>T_c$, the above result for $\uuline{I}{}_n$ agrees with the
collision integral derived by Jeon and Mullin \cite{Jeon1988a}, and thus
our kinetic theory above $T_c$ reduces to the Jeon-Mullin
theory. Furthermore, if one considers temperatures well above $T_c$
where the Bose enhancement of collisions can be neglected, one
recovers the kinetic equation for a classical spin-1/2 gas introduced by 
Lhuillier and Lalo\"e~\cite{Lhuillier1982b} (we do not include the spin rotation term in the
collision integral, as we discuss in Appendix \ref{sec:collisions}).

If we set the internal coherence of the noncondensate $W_{12}=0$ to
zero, and take the coupling field to zero $\vec \Omega = 0$, then our
theory becomes equivalent to that for a binary mixture of two different atomic species
(with the same mass). 
For the case where the population of one of the components is zero, our
kinetic theory reduces to the ZNG kinetic theory for the
single-component Bose-condensed gas \cite{Zaremba1999a}.

In this paper we are interested in the general case where the internal
coherence of the noncondensate is finite. Comparing
Eqs.~(\ref{spinor_GP}) and (\ref{KE}) with (3) and (4) of Jackson and
Zaremba~\cite{Jackson2002a}, we see that the structure of the spin-1/2 kinetic
theory is the same as the ZNG theory for a single component gas. In
particular, the collision integrals $\uuline{I}{}_c$ and
$\uuline{I}{}_n$ are generalizations of the $C_{12}$ and $C_{22}$
collision integrals given by Eqs.~(23a) and (23b) in ZNG \cite{Zaremba1999a}.
The spin-1/2 Bose-condensed gas, however, is a much richer system due to the extra
spin degrees of freedom. For a given excitation symmetry, the number
of collective modes for the spin-1/2 gas is three times as large
compared to the single component gas, for one has total density,
relative density (or longitudinal spin), and transverse spin dynamics
to consider. Similarly to the single component gas, one must also
consider the relative motion of the condensate and noncondensate for
each of these degrees of freedom. In order to better understand the
various types of motion, it is extremely useful to recast the theory
into a form that separates out the spin and total density degrees
explicitly.

\section{Spin hydrodynamics}
\label{sec:hydro}
The condensate spinor order parameter can be described in terms of
density and spin variables, which obey equations of motion exactly
equivalent to the spinor GP equation. The condensate spinor
can be written as \cite{Matthews1999a} 
\begin{equation}
\underline\Phi({\bf r},t)=
\left(\matrix{ \Phi_1({\bf r},t) \cr
\Phi_2({\bf r},t) \cr}\right)
=\sqrt{n_c({\bf r},t)}e^{i\alpha_c({\bf r},t)}
\left(\matrix{ e^{-i\phi_c({\bf r},t)/2}
\cos\frac{\theta_c({\bf r},t)}{2}
\cr e^{i\phi_c({\bf r},t)/2}
\sin\frac{\theta_c({\bf r},t)}{2}
\cr }\right),
\label{cond_spinor}
\end{equation}
where $n_c=n_{c1}+n_{c2}=|\Phi_1|^2+|\Phi_2|^2$ is the total
condensate density, and $\alpha_c$ is the overall phase of the
condensate wavefunction. 
The two phase angles $\phi_c$ and $\theta_c$ define the spin (Bloch vector) of the condensate
\begin{equation}
\vec{S}_c({\bf r},t)={\rm
Tr}\{\vec{\uuline{\sigma}}~\uuline{n}{}_c({\bf r},t)\} \equiv
n_c({\bf r},t)\vec{M}_c({\bf r},t),
\end{equation} 
where $\vec{M}_c$ is the reduced condensate spin vector
\begin{equation}
\vec{M}_c=(\cos\phi_c\sin\theta_c,\,\sin\phi_c\sin\theta_c,\,\cos\theta_c).
\end{equation}
The local velocity flow ${\bf v}_c({\bf r},t)$ for the total condensate density $n_c({\bf r},t)$ 
is given by (see also Eq.~(\ref{vc_def}))
\begin{equation}
{\bf v}_c
=\frac{\hbar}{m}\left({\boldsymbol\nabla}\alpha_c-\frac{1}{2}\cos\theta_c
{\boldsymbol\nabla}\phi_c\right).
\label{vc_def2}
\end{equation}
We remark that the velocity field of a spin-1/2 order parameter is not irrotational (i.e. $\boldsymbol\nabla \times
{\bf v}_c \neq 0$) in general, a well known result.
Finally, the local energy of the condensate is given by
(see also Eq.~(\ref{ec_def}))
\begin{equation}
\varepsilon_c({\bf r},t) \equiv \mu_c({\bf
r},t)+\frac{mv_c^2({\bf r},t)}{2}, 
\label{ec_def2}
\end{equation}
where the explicit formula of the local chemical potential for the condensate is
given from Eq.~(\ref{eq_muc}):
\begin{equation}
\mu_c({\bf r},t)\equiv 
-\frac{\hbar^2\nabla^2\sqrt{n_c({\bf r},t)}}
{2m\sqrt{n_c({\bf r},t)}}+U_c({\bf r},t)
+\frac{\hbar}{2}\vec{\Omega}_c ({\bf r},t)\cdot\vec{M}_c({\bf r},t)
+\frac{\hbar^2}{8m}[\boldsymbol\nabla\vec{M}_c({\bf r},t)]^2.
\label{muc_def}
\end{equation}

The equations of motion for the condensate total density, velocity,
and spin density can be obtained directly from the spinor GP equation
(\ref{spinor_GP}) using Eq.~(\ref{cond_spinor})
\begin{equation}
\frac{\partial n_c}{\partial t}+{\boldsymbol\nabla}\cdot(n_c{\bf v}_c)
=\left.\frac{\partial n_c}{\partial t}\right|_{\rm coll},
\label{eq_nc}
\end{equation}
\begin{eqnarray}
m\frac{\partial v_{c\mu}}{\partial t}&=&
-\frac{\partial \varepsilon_c}{\partial x_{\mu}}+
\frac{\hbar}{2}\vec{\Omega}_c\cdot\frac{\partial \vec{M}_c}{\partial x_{\mu}}
+\frac{\hbar}{2}v_{c\nu}
\frac{\partial \vec{M}_c}{\partial x_{\nu}}
\cdot\left(\vec{M}_c\times\frac{\partial \vec{M}_c}{\partial x_{\mu}}\right) \cr
&&-\frac{\hbar^2}{4mn_c}\frac{\partial\vec{M}_c}{\partial x_{\mu}}
\cdot\frac{\partial}{\partial x_{\nu}}\left( n_c \frac{\partial\vec{M}_c}
{\partial x_{\nu}} \right)
+\frac{1}{2}\left(\vec{M}_c\times\frac{\partial\vec{M}_c}
{\partial x_{\mu}}\right)\cdot{\rm Tr}(\uuline{\vec{\sigma}}
~ \uuline{R}),
\end{eqnarray}
\begin{equation}
\frac{\partial \vec{S}_c}{\partial t}+\frac{1}{m}{\boldsymbol\nabla}\cdot\vec{\bf J}_c
=\vec{\Omega}_c'\times\vec{S}_c+\frac{g_{12}}{\hbar}\vec{S}_n\times\vec{S}_c
+\left.\frac{\partial\vec{S}_c}{\partial t}\right|_{\rm coll},
\label{eq_sc}
\end{equation}
where $\mu$ and $\nu$ in Eq.~(76) are Cartesian components and repeated
subscripts are summed.
Here the noncondensate spin is defined as
\begin{equation}
\vec{S}_n({\bf r},t)\equiv {\rm Tr}\{\uuline{\vec{\sigma}}~
\uuline{\tilde n}({\bf r},t)\}.
\end{equation}
The condensate spin current is given explicitly in terms of the other
condensate variables by
\begin{equation}
\vec{{\bf J}}_c\equiv m\vec{S}_c{\bf v}_c-\frac{\hbar}{2n_c}\vec{S}_c\times
{\boldsymbol\nabla}\vec{S}_c.
\label{Jc}
\end{equation}
We note that $\vec{{\bf J}}_c({\bf r},t)$ is a tensor field, and the operation ${\boldsymbol\nabla}\cdot\vec{\bf J}_c({\bf r},t)$ should be interpreted as
\begin{equation}
{\boldsymbol\nabla}\cdot\vec{\bf J}_c({\bf r},t) = {\partial \over {\partial x}} \vec{J}_{cx}({\bf r},t) + {\partial \over {\partial y}} \vec{J}_{cy}({\bf r},t)+{\partial \over {\partial z}} \vec{J}_{cz}({\bf r},t).
\end{equation}
The collisional contributions in Eqs.~(\ref{eq_nc}) and (\ref{eq_sc}) are given by 
\begin{equation}
\left.\frac{\partial n_c}{\partial t}\right|_{\rm coll}
=-\frac{1}{\hbar}{\rm Tr}\{\uuline{n}{}_c,\uuline{R}\},
\end{equation}
\begin{equation}
\left.\frac{\partial \vec{S}_c}{\partial t}\right|_{\rm coll}
=-\frac{1}{\hbar}{\rm Tr}[\vec{\uuline{\sigma}}\{\uuline{n}{}_c,\uuline{R}\}].
\end{equation}

We now turn our attention to the noncondensate kinetic equation.
The noncondensate hydrodynamic equations can be obtained by taking moments of the
kinetic equation, and inserting a local equilibrium form for $\uuline{W}({\bf r},{\bf p},t)$.
The first three moment equations define the spin density, spin current, and
spin pressure of the noncondensate component:
\begin{equation}
\vec{S}_n({\bf r},t)\equiv
\int\frac{d{\bf p}}{(2\pi\hbar)^3}
{\rm Tr}\vec{\uuline{\sigma}}~
\uuline{W}({\bf r},{\bf p},t),
\end{equation}
\begin{equation}
\vec{\bf J}_n({\bf r},t)\equiv
\int\frac{d{\bf p}}{(2\pi\hbar)^3}{\bf p}
{\rm Tr}\vec{\uuline{\sigma}}~
\uuline{W}({\bf r},{\bf p},t),
\end{equation}
\begin{equation}
\vec{P}_{\mu\nu}({\bf r},t)\equiv\int\frac{d{\bf p}}{(2\pi\hbar)^3}
\frac{p_{\mu}p_{\nu}}{m}{\rm Tr}\vec{\uuline{\sigma}}~
\uuline{W}({\bf r},{\bf p},t),
\end{equation}
It is straightforward to derive the general moment equations from the
kinetic equation Eq.~(\ref{KE}):
\begin{equation}
\frac{\partial \vec{S}_n}{\partial t}
+\frac{1}{m}{\boldsymbol\nabla}\cdot\vec{{\bf J}}_n
=\vec{\Omega}_n'\times\vec{S}_n+\frac{g_{12}}{\hbar}
\vec{S}_c\times\vec{S}_n+
\left.\frac{\partial \vec{S}_n}{\partial t}\right|_{\rm coll},
\label{eq_Sn}
\end{equation}
\begin{equation}
\frac{\partial\vec{J}_{n\mu}}{\partial t}+\frac{\partial \vec{P}_{\mu\nu}}
{\partial x_{\nu}}+\frac{\partial U_n}{\partial x_{\mu}}\vec{S}_n
-\vec{\Omega}_n\times\vec{J}_{n\mu}-\frac{\hbar}{2}\tilde{n}
\frac{\partial\vec{\Omega}_n}{\partial x_{\mu}}
=\left. \frac{\partial \vec{J}_{n\mu}} {\partial t} \right|_{\rm coll}.
\label{eq_Jn}
\end{equation}
The subscript $\mu$ in $\vec{J}_{n\mu}$ signifies a coordinate space vector component of
$\vec{\bf{J}}_n$. 
Here the collisional contributions are given by
\begin{equation}
\left.\frac{\partial \vec{S}_n}{\partial t}\right|_{\rm coll}
\equiv 
\int\frac{d{\bf p}}{(2\pi\hbar)^3}
{\rm Tr}\uuline{\vec{\sigma}}
\left(\uuline{I}{}_c[\uuline{W}]+\uuline{I}{}_n[\uuline{W}]\right),
\label{Sn_coll}
\end{equation}
\begin{equation}
\left.\frac{\partial \vec{\bf{J}}_n}{\partial t}\right|_{\rm coll}
\equiv 
\int\frac{d{\bf p}}{(2\pi\hbar)^3}
{\bf p}{\rm Tr}\uuline{\vec{\sigma}}
\left(\uuline{I}{}_c[\uuline{W}]
+\uuline{I}{}_n[\uuline{W}]\right).
\end{equation}

We note that the equation for $\vec{S}_n$ in Eq.~(\ref{eq_Sn}) involves the collisional
contribution described by the following collision matrices
\begin{equation}
\uuline{\Gamma}{}_{n} \equiv \int \frac{d{\bf p}}
{(2\pi\hbar)^3}\uuline{I}{}_n[\uuline{W}], 
\label{Gamman_def}
\end{equation}
\begin{equation}
\uuline{\Gamma}{}_{c} \equiv \int \frac{d{\bf p}}
{(2\pi\hbar)^3}\uuline{I}{}_c[\uuline{W}].
\label{Gammac_def}
\end{equation}
Eq.~(\ref{Sn_coll}) can be written in terms of these collisional matrices as 
$\partial \vec S_n / \partial t |_{\rm{coll}} =
\rm{Tr}\uuline{\vec{\sigma}}(\uuline{\Gamma}{}_c + \uuline{\Gamma}{}_n)$.
In the case of equal scattering lengths we are considering here, one immediately finds
$\uuline{\Gamma}{}_n=0$, i.e., collisions
between noncondensate atoms conserve the density and spin density.
In contrast, collisions between the condensate and noncondensate do
not conserve the density and spin density of each component since
atoms can scatter into and out of the condensate. 
One can show that the dissipative term $i\uuline{R}$ appearing in the
GP equation is related to $\uuline{\Gamma}{}_c$, and thus to $\uuline{I}{}_c$, according to
\begin{equation}
\uuline{\Gamma}{}_c\equiv \frac{1}{\hbar}\{\uuline{n}{}_c,\uuline{R}\}
=-\left.\frac{\partial \uuline{n}{}_c}{\partial t}\right|_{\rm coll},
\label{Gamma_c_R}
\end{equation}
where $\partial \uuline{n}{}_c/\partial t|_{\rm coll}$ is the collisional
contribution to $\partial \uuline{n}{}_c/\partial t$.
The relationship between $\uuline{I}{}_c$ and $\uuline{\Gamma}{}_c$ in
Eq.~(\ref{Gammac_def}), along with Eq.~(\ref{Gamma_c_R}), ensures the conservation of total spin and
atom number.
The collision matrix $\uuline{\Gamma}{}_c$, which is a source of damping of density and spin
excitations in the condensate, vanishes only when the
condensate and noncondensate are in complete local equilibrium with
each other, as we will show later.
When we consider unequal scattering lengths e.g., $g_{11}\neq g_{12}$, $\uuline{\Gamma}{}_n$ also makes
a finite contribution to Eq.~(\ref{eq_Sn}), giving rise to relaxation of the transverse spin
component. 
We discuss this in Section \ref{sec:different_g}.

In order to reduce the general moment equations to a closed set of equations,
we assume that rapid $\uuline{I}{}_n$ collisions
brings the noncondensate into local equilibrium. 
To be in this collision-dominated hydrodynamic regime, the collective mode of frequency
$\omega$ must satisfy $\omega\tau<1$, where $\tau$ is some appropriate relaxation time for reaching
local equilibrium.
One can show that the appropriate local equilibrium distribution
function is given by \cite{Jeon1988a}
\begin{equation}
\uuline{W}^{\rm leq}({\bf r},{\bf p},t)=\sum_s\frac{1}{2}[\uuline{1}
+s\vec{e}_n({\bf r},t)\cdot\uuline{\vec{\sigma}}]f_s({\bf r},{\bf p},t),
\label{W_local}
\end{equation}
where $s$ is $+1$ or $-1$ and $\vec{e}_n$ is a unit vector representing the local direction of
the noncondensate spin moment.
One can express $\vec{e}_n$ using the transverse and longitudinal angles of the spin
vector as
\begin{equation}
\vec{e}_n=(\cos\phi_n\sin\theta_n,\sin\phi_n\sin\theta_n,\cos\theta_n).
\end{equation}
The distributions for the dressed states (defined along the local polarization axis $\vec e_n$)
$f_\pm({\bf r},{\bf p},t)$ are given by 
\begin{equation}
f_{s}=\left(\exp\left\{\beta\left[\frac{({\bf p}-m{\bf v}_n)^2}{2m}+U_n+\frac{\hbar}{2}
\vec{\Omega}_n\cdot\vec{M}_n-\tilde\mu_s\right]\right\}-1\right)^{-1}.
\label{f_pm}
\end{equation}
Here $\tilde\mu_s({\bf r}, t)$ is a chemical potential for a dressed state $s$, which may vary in
position and time. 
With this local equilibrium distribution, the local spin density is given by
\begin{equation}
\vec{S}_n=\vec{e}_n(\tilde{n}_+-\tilde{n}_-)\equiv\vec{e}_n\tilde{n}M_n,~
M_n\equiv(\tilde n_+-\tilde n_-)/(\tilde n_++\tilde n_-),
\label{Sn_le}
\end{equation}
where
\begin{equation}
\tilde{n}_s=\frac{1}{\lambda_T^3}g_{3/2}(z_s),~~
z_s=\exp\left[\beta\left(\tilde\mu_s-U_n-
\frac{\hbar}{2}\vec{\Omega}_n\cdot\vec{M}_n\right)\right],
\label{ntilde_le}
\end{equation}
and $\lambda_T\equiv(2\pi\hbar^2/mk_{\rm B}T)^{1/2}$ is the
thermal de Broglie wavelength. 
For later use, we also express the condensate component in the analogous form
\begin{equation}
\uuline{n}{}_c({\bf r},t)=\frac{n_c({\bf r},t)}{2}[\uuline{1}+\vec{e}_c({\bf r},t)
\cdot\vec{\uuline{\sigma}}].
\end{equation}
We note that the condensate spin is always fully polarized with the local
direction $\vec{e}_c$, and thus one always has $\vec{e}_c=\vec{M}_c$.

It is straightforward to verify that this local equilibrium distribution satisfies
$\uuline{I}{}_n[\uuline{W}^{\rm leq}]=0$,
independent of $\tilde\mu_s$ and $\vec{e}_n$
(this condition indeed defines the local equilibrium solution).
In contrast, $\uuline{I}{}_c$ in general does not vanish, i.e.,
\begin{eqnarray}
\uuline{I}{}_c[\uuline{W}^{\rm leq}]&=&
\frac{\pi g^2 n_c}{2\hbar} \int\frac{d{\bf p}_1}{(2\pi\hbar)^3}
\int d{\bf p}_2\int d{\bf p}_3 \cr
&&\times\delta(m{\bf v}_c+{\bf p}_1-{\bf p}_2-{\bf p}_3)
\delta(\varepsilon_c+\tilde\varepsilon_{p_1}-
\tilde\varepsilon_{p_2}-\tilde\varepsilon_{p_3}) \cr
&&\times\sum_{s,s'=\pm1}\left\{1-e^{-\beta[\tilde\mu_{s'}-
m({\bf v}_n-{\bf v}_c)^2/2-\mu_c]}\right\} \cr
&&\times\{[\delta({\bf p}-{\bf p}_1)-\delta({\bf p}-{\bf p}_2)]
(\uuline{1}+s\vec{e}_n\cdot\vec{\uuline{\sigma}})(1+s'\vec{e}_n\cdot\vec{e}_c)
(1+\delta_{s,s'}) \cr
&&\times[\delta({\bf p}-{\bf p}_1)-\delta({\bf p}-{\bf p}_3)](1-\delta_{s,s'})
\vec{\uuline{\sigma}}\cdot[\vec{e}_c-\vec{e}_n(\vec{e}_n\cdot\vec{e}_c)] \cr
&&-\delta({\bf p}-{\bf p}_3)
[(1+s'\vec{e}_n\cdot\vec{e}_c)\uuline{1}+(\vec{e}_c+s'\vec{e}_n)\cdot
\vec{\uuline{\sigma}}](1+\delta_{s,s'})\}\cr
&&\times(1+f_{1s})f_{2s}f_{3s'}
\label{IcWleq}
\end{eqnarray}
This equation is analogous to Eq.~(40) of ZNG~\cite{Zaremba1999a} for the $C_{12}$
collision integral in a single-component Bose gas.
If we impose the condition that the first square bracket vanish for both
components $s=\pm 1$, one must have $\tilde\mu_+=\tilde\mu_-=\mu_c+m({\bf v}_n-{\bf v}_c)^2/2$.
This condition is too restrictive, since it results in $f_+=f_-$, which occurs
when the polarization of the noncondensate vanishes.
In order to satisfy $\uuline I{}_c=0$ while allowing for a {\it{finite}} noncondensate spin moment, 
the condensate and noncondensate local spin polarizations must be parallel $\vec{e}_n=\vec{e}_c$ 
(or antiparallel $\vec{e}_n=-\vec{e}_c$). 
In this case, one can show that all the contributions from the $f_-$ (or $f_+$ in the antiparallel
case) component vanish.
Then local equilibrium between the condensate and noncondensate is specified by the conditions 
(for the parallel case) 
\begin{equation}
\vec{e}_n=\vec{e}_c,
\label{enc_local}
\end{equation}
\begin{equation}
\tilde\mu_+=\mu_c+\frac{m}{2}({\bf v}_n-{\bf v}_c)^2.
\label{munc_local}
\end{equation}
The role of the condition for the chemical potential in Eq.~(\ref{munc_local}) has been
discussed extensively for the single-component Bose gas \cite{Zaremba1999a,Williams2001b};
for example, collisional equilibration between the condensate and noncondensate gives rise
to damping of condensate collective modes \cite{Williams1999c}.

In the spin-1/2 case, there is an additional condition in Eq.~(\ref{enc_local}) for the
the spin orientations. This additional condition means that there is a collisional relaxation process 
related to the relative orientation of the condensate and noncondensate spins, which will
give rise to damping of condensate spin dynamics.
This relaxation process comes into the dynamical equation of motion for the condensate
and noncondensate spins [see Eqs.~(\ref{Sn_coll}) and (\ref{Gamma_c_R})] through
\begin{eqnarray}
\uuline{\Gamma}{}_c[\uuline{W}^{\rm leq}]&=&
-\frac{\pi g^2 n_c}{2\hbar} \int\frac{d{\bf p}_1}{(2\pi\hbar)^3}
\int \frac{d{\bf p}_2}{(2\pi\hbar)^3}\int d{\bf p}_3 \cr
&&\times\delta(m{\bf v}_c+{\bf p}_1-{\bf p}_2-{\bf p}_3)
\delta(\varepsilon_c+\tilde\varepsilon_{p_1}-
\tilde\varepsilon_{p_2}-\tilde\varepsilon_{p_3}) \cr
&&\times\sum_{s,s'=\pm1}\left\{1-e^{-\beta[\tilde\mu_{s'}-
m({\bf v}_n-{\bf v}_c)^2/2-\mu_c]}\right\} \cr
&&[(1+s'\vec{e}_n\cdot\vec{e}_c)\uuline{1}+(\vec{e}_c+s'\vec{e}_n)\cdot
\vec{\uuline{\sigma}}](1+\delta_{s,s'})\cr
&&\times(1+f_{1s})f_{2s}f_{3s'}.
\label{GammacWleq}
\end{eqnarray}
The role of this term in the spin dynamics will be more explicitly seen later.

To obtain the equation for the spin current, we must include the
deviation from local equilibrium.
For simplicity, we consider the situation where the condensate and
noncondensate components have no particle currents.
We then follow the linearization procedure given by Jeon and Mullin~\cite{Jeon1988a},
expanding the distribution function around local equilibrium as
\begin{equation}
\uuline{W}=\uuline{W}^{\rm leq}+\delta\uuline{W}.
\end{equation}
The lowest order correction $\delta \uuline{W}$ is found to 
take the following variational form \cite{Jeon1988a}:
\begin{equation}
\delta\uuline{W}=\frac{1}{2}\sum_{s}(\uuline{1}+s\vec{e}_n\cdot\vec{\uuline{\sigma}})
f_s(1+f_s)\psi^{\parallel}_s+\frac{1}{2}\vec{\psi}^{\perp}\cdot\vec{\uuline{\sigma}}
\sum_s sf_s,
\end{equation}
where $\vec{\psi}^{\perp}$ is perpendicular to $\vec{e}_n$, i.e.,
$\vec{\psi}^{\perp}\cdot\vec{e}_n=0$.
The functions $\psi_s^{\parallel}$ and $\vec{\psi}^{\perp}$ are associated with the 
spin current through
\begin{equation}
\psi^{\parallel}_s({\bf p})=s\frac{\beta}{2mn_s}{\bf J}_n^{\parallel}\cdot{\bf p},~~
\vec{\psi}^{\perp}({\bf p})=\frac{1}{m\sum_s s\tilde P_s}\vec{\bf J}_n^{\perp}\cdot
{\bf p},
\label{psi_correction}
\end{equation}
where ${\bf J}_n^{\parallel}$ and $\vec{\bf J}_n^{\perp}$ denote the
longitudinal and transverse components 
\begin{equation}
\vec{\bf J}_n=\vec{\bf J}_n^{\parallel}+\vec{\bf J}_n^{\perp}
=\vec{e}_n{\bf J}_n^{\parallel}+\vec{\bf J}_n^{\perp},~~
\vec{\bf J}_n^{\perp}\perp\vec{e}_n,
\end{equation}
and $\tilde P_s$ is defined by
\begin{equation}
\tilde P_s\equiv\int\frac{d{\bf p}}{(2\pi\hbar)^3}\frac{p^2}{3m}f_s
=\frac{k_{\rm B}T}{\lambda_T^3}g_{5/2}(z_s).
\label{Ptilde}
\end{equation}
With this approximate form of the distribution function, the spin
pressure tensor becomes diagonal:
\begin{equation}
\vec{P}_{\mu\nu}=\vec{e}_n\delta_{\mu\nu}\sum_s s\tilde P_s
\equiv\delta_{\mu\nu}\vec{P},~~
\vec{P}\equiv\vec{e}_n\sum_s s\tilde P_s.
\label{Pmunu}
\end{equation}
We note that $\vec{P}$ and $\vec{S}_n$ are related through Eqs.~(\ref{Sn_le}), 
(\ref{ntilde_le}), (\ref{Ptilde}), and (\ref{Pmunu}), along with the condition
$\tilde n_1({\bf r},t)+\tilde n_2({\bf r},t)=\tilde n_0({\bf r})$.
While Jeon and Mullin \cite{Jeon1988a} found this variational solution for
a noncondensed gas above $T_c$, it is also applicable to the thermal gas
in the Bose-condensed phase, because our kinetic equation in Eq.~(\ref{KE}) 
is formally the same as that of Ref.~\cite{Jeon1988a} (i.e., the condensate contributions are
buried in the effective potential $\uuline{U}$ and the collision integral). 
We then obtain the following equation for the noncondensate spin current:
\begin{equation}
\frac{\partial \vec{\bf J}_n}{\partial t}
+{\boldsymbol\nabla}\vec{P}+{\boldsymbol\nabla}U_n\vec{S}_n+\frac{\hbar}{2}
{\boldsymbol\nabla}\vec{\Omega}_n\tilde n-\vec{\Omega}_n\times\vec{\bf J}_n=
\left. \frac{\partial \vec{\bf J}_n}{\partial t}\right|_{\rm coll}.
\end{equation}

To evaluate the collisional contributions in the hydrodynamic equations,
we must linearize the collision integrals in $\psi^{\parallel}_s$ and
$\vec{\psi}^{\perp}$:
\begin{equation}
\uuline{I}{}_c[\uuline{W}]=\uuline{I}{}_c[\uuline{W}{}^{\rm leq}]
+\delta\uuline{I}{}_c[\delta\uuline{W}],
\label{linear_Ic}
\end{equation}
\begin{equation}
\uuline{I}{}_n[\uuline{W}]=\delta\uuline{I}{}_n[\delta\uuline{W}].
\label{linear_In}
\end{equation}
The contribution from $\uuline{W}^{\rm leq}$ in Eq.~(\ref{linear_Ic}) leads to
$\uuline{\Gamma}{}_c$ given in Eq.~(\ref{GammacWleq}).
In \ref{linear_collision}, we give detailed calculations of the linearized collision
integrals.
Here we simply present the main results.
Assuming that the condensate spin and noncondensate spin are almost in local equilibrium
with each other, we linearize Eq.~(\ref{GammacWleq}) in $\mu_{\rm diff}\equiv 
\tilde \mu_+-\mu_c$ and $\vec{e}_n-\vec{e}_c$ to obtain
\begin{eqnarray}
\left.\frac{\partial\vec{S}_n}{\partial t}\right|_{\rm coll}
&=&{\rm Tr}\left(\vec{\uuline{\sigma}}\,\uuline{\Gamma}{}_c[\hat{W}^{\rm leq}]\right) \cr
&=&-\frac{\tilde n\beta\mu_{\rm diff}}{\tilde\tau_c^{\parallel}}
\left(\frac{\vec{e}_n+\vec{e}_c}{2}\right)
-\frac{\tilde n}{\tilde\tau_c^{\perp}}(\vec{e}_n-\vec{e}_c).
\label{Sn_coll2}
\end{eqnarray}
We also obtain
\begin{equation}
\left.\frac{\partial\vec{\bf J}_n}{\partial t}\right|_{\rm coll}
=\int\frac{d{\bf p}}{(2\pi\hbar)^3}{\bf p}{\rm Tr}\left[\vec{\uuline{\sigma}}
(\delta\uuline{I}{}_n+\delta\uuline{I}{}_c)\right]
=-\frac{\vec{{\bf J}}_n^{\parallel}}{\tau_D^{\parallel}}
-\frac{\vec{{\bf J}}_n^{\perp}}{\tau_D^{\perp}}.
\end{equation}
Explicit formulas of the four relaxation times 
$\tilde\tau_c^{\parallel}$, $\tilde\tau_c^{\perp}$, $\tau_D^{\parallel}$, and
$\tau_D^{\perp}$ are given in Appendix \ref{linear_collision}. 
Here we note that the spin diffusion relaxation times $\tau_D^{\parallel}$ and
$\tau_D^{\perp}$ involve the contribution from $\uuline{I}{}_c$ term,
which represents collisions between the condensate and noncondensate.

In summary, we have obtained the following coupled spin hydrodynamic
equations, assuming no overall mass 
currents for the condensate and noncondensate:
\begin{eqnarray}
\frac{\partial\vec{S}_c}{\partial t}+\frac{1}{m}{\boldsymbol\nabla}\cdot\vec{\bf J}_c&=&
\vec{\Omega}'_c\times\vec{S}_c+\frac{g_{12}}{\hbar}\vec{S}_n\times\vec{S}_c
\cr
&&+\frac{n_c\beta\mu_{\rm diff}}{\tau_c^{\parallel}}
\left(\frac{\vec{e}_n+\vec{e}_c}{2}\right)-\frac{n_c}{\tau_c^{\perp}}
(\vec{e}_c-\vec{e}_n),
\label{spinhydro1}
\end{eqnarray}
\begin{equation}
\vec{\bf J}_c=-\frac{\hbar}{2n_c}\vec{S}_c\times{\boldsymbol\nabla}\vec{S}_c,
\label{spinhydro2}
\end{equation}
\begin{eqnarray}
\frac{\partial\vec{S}_n}{\partial t}+\frac{1}{m}{\boldsymbol\nabla}\cdot\vec{\bf J}_n
&=&\vec{\Omega}'_n\times\vec{S}_n+\frac{g_{12}}{\hbar}\vec{S}_c\times\vec{S}_n
\cr
&&-\frac{\tilde n\beta\mu_{\rm diff}}{\tilde\tau_c^{\parallel}}
\left(\frac{\vec{e}_n+\vec{e}_c}{2}\right)-\frac{\tilde n}{\tilde\tau_c^{\perp}}
(\vec{e}_n-\vec{e}_c),
\label{spinhydro3}
\end{eqnarray}
\begin{equation}
\frac{\partial \vec{\bf J}_n}{\partial t}
+{\boldsymbol\nabla}\vec{P}+{\boldsymbol\nabla}U_n\vec{S}_n+\frac{\hbar}{2}
{\boldsymbol\nabla}\vec{\Omega}_n\tilde n-\vec{\Omega}_n\times\vec{\bf J}_n=
-\frac{\vec{{\bf J}}_n^{\parallel}}{\tau_D^{\parallel}}
-\frac{\vec{{\bf J}}_n^{\perp}}{\tau_D^{\perp}}.
\label{spinhydro4}
\end{equation}
The modified external field can be written as
\begin{equation}
\vec{\Omega}'_c=\vec{\Omega}+\hat{z}\frac{1}{2\hbar}
[(g_{11}-g_{22})(n_c+2\tilde n)+(g_{11}+g_{22}-2g_{12})(S_c^z+2S_n^z)],
\end{equation}
\begin{equation}
\vec{\Omega}'_n=\vec{\Omega}+\hat{z}\frac{1}{\hbar}
[(g_{11}-g_{22})(n_c+\tilde n)+(g_{11}+g_{22}-2g_{12})(S_c+S_n^z)],
\end{equation}
while the effective field for the noncondensate is
\begin{equation}
\vec{\Omega}_n=\vec{\Omega}'_n+\frac{g_{12}}{\hbar}(\vec{S}_c+\vec{S}_n).
\end{equation}

Here we remark on some important features of the spin hydrodynamic equations.
\begin{enumerate}
\item The condensate spin current involves the spatial gradient of the condensate
spin density itself. This term exert a torque on the condensate spins, which causes
spin untwisting observed in the JILA experiment~\cite{Matthews1999a}. It is important to note that this
term dose not involve the interaction, and is a sort of quantum pressure term (also 
referred to as the "quantum torque"~\cite{Holland1995a}).
This is in sharp contrast with the noncondensate spin dynamics. The exchange mean field
plays a crucial role in the noncondensate spin current, which causes collective spin
waves.
\item The above equations involve coupling between the condensate and noncondensate spins 
through the exchange mean-field terms, such as $(g_{12}/\hbar)\vec{S}_c\times\vec{S}_n$, which
exerts a mutual torque when the condensate and noncondensate spins are not aligned.
\item The two components are also coupled through the collisional relaxation terms.
These terms try to bring the two components into complete local equilibrium
i.e., $\mu_+=\mu_c$ and $\vec{e}_n=\vec{e}_c$.
These conditions determine the length and the direction of the spins.
\end{enumerate}

\section{Effect of different scattering lengths}
\label{sec:different_g}
The collision integrals given in Eqs.~(\ref{R_symmetric}-\ref{Ic_symmetric}) were obtained for the
case of equal scattering lengths $g_{ij}\equiv g$, corresponding to SU(2)-symmetric interactions. 
For this special case, collisions make no explicit contribution to the decay of the net transverse
spin. In this section we consider the effect of unequal scattering lengths,
which turn out to contribute to the decay of the net transverse spin.

We first consider the role of $\uuline{I}{}_n$ integral.
From Eq.~(\ref{Gamman_def}), the collision matrix $\uuline{\Gamma}{}_n$ is given by
\begin{eqnarray}
\uuline{\Gamma}{}_{n}&=&\int\frac{d{\bf p}}{(2\pi\hbar)^3}
\uuline{I}{}_{n}[\uuline{W}] \cr
&=&\frac{\pi}{\hbar}\int\frac{d{\bf p}_1}{(2\pi\hbar)^3}
\int\frac{d{\bf p}_2}{(2\pi\hbar)^3}\int\frac{d{\bf p}_3}{(2\pi\hbar)^3}
\int d{\bf p}_4 \cr
&&\times\delta(\tilde\varepsilon_{p_1}+\tilde\varepsilon_{p_2}
-\tilde\varepsilon_{p_3}-\tilde\varepsilon_{p_4})
\delta({\bf p}_1+{\bf p}_2-{\bf p}_3-{\bf p}_4) \cr
&&\times\bigl[\{\uuline{1}+\uuline{W}({\bf p}_1),\uuline A^{<}({\bf p}_2;{\bf p}_3,{\bf p}_4)\}\cr
&&-\{\uuline{W}({\bf p}_1),\uuline A^{>}({\bf p}_2;{\bf p}_3,{\bf p}_4)\}\bigr],
\end{eqnarray}
where the explicit formulas for the matrices $A^{\stackrel{<}{>}}$ are given
by Eq.~(\ref{A<}).
After some rearrangement, we obtain
\begin{eqnarray}
\Gamma_{nij}&=&\frac{\pi}{\hbar}\int\frac{d{\bf p}_1}{(2\pi\hbar)^3}
\int\frac{d{\bf p}_2}{(2\pi\hbar)^3}\int\frac{d{\bf p}_3}{(2\pi\hbar)^3}
\int d{\bf p}_4 \cr
&&\times\delta(\tilde\varepsilon_{p_1}+\tilde\varepsilon_{p_2}
-\tilde\varepsilon_{p_3}-\tilde\varepsilon_{p_4})
\delta({\bf p}_1+{\bf p}_2-{\bf p}_3-{\bf p}_4) \cr
&&\times\Bigl\{
\sum_{klm}(g_{kj}-g_{ik})g_{ml}\bigl[
{W}^{>}_{im}({\bf p}_1)W^{>}_{kl}({\bf p}_2)W^{<}_{lk}({\bf p}_3)
W^{<}_{mj}({\bf p}_4) \cr
&&-{W}^{<}_{im}({\bf p}_1)W^{<}_{kl}({\bf p}_2)W^{>}_{lk}({\bf p}_3)
W^{>}_{mj}({\bf p}_4) \cr
&&+
{W}^{>}_{im}({\bf p}_1)W^{>}_{kl}({\bf p}_2)W^{<}_{mk}({\bf p}_3)
W^{<}_{lj}({\bf p}_4) \cr
&&-{W}^{<}_{im}({\bf p}_1)W^{<}_{kl}({\bf p}_2)W^{>}_{mk}({\bf p}_3)
W^{>}_{lj}({\bf p}_4)\bigr] \Bigr\}
\end{eqnarray}
One immediately finds that $\Gamma_{nii}=0$ regardless of the form
of $\uuline W({\bf p})$.
This means that the collisions between noncondensate atoms conserve
the number of noncondensate atoms in each hyperfine state.
We now focus on the finite contribution of the transverse component, namely
\begin{eqnarray}
\Gamma_{n12}&=&\frac{\pi}{\hbar}\int\frac{d{\bf p}_1}{(2\pi\hbar)^3}
\int\frac{d{\bf p}_2}{(2\pi\hbar)^3}\int\frac{d{\bf p}_3}{(2\pi\hbar)^3}
\int d{\bf p}_4 \cr
&&\times\delta(\tilde\varepsilon_{p_1}+\tilde\varepsilon_{p_2}
-\tilde\varepsilon_{p_3}-\tilde\varepsilon_{p_4})
\delta({\bf p}_1+{\bf p}_2-{\bf p}_3-{\bf p}_4) \cr
&&\times\{
\sum_{klm}(g_{k2}-g_{1k})g_{ml}[
{W}^{>}_{1m}({\bf p}_1)W^{>}_{kl}({\bf p}_2)W^{<}_{lk}({\bf p}_3)
W^{<}_{m2}({\bf p}_4) \cr
&&-{W}^{<}_{1m}({\bf p}_1)W^{<}_{kl}({\bf p}_2)W^{>}_{lk}({\bf p}_3)
W^{>}_{m2}({\bf p}_4) \cr
&&+
{W}^{>}_{1m}({\bf p}_1)W^{>}_{kl}({\bf p}_2)W^{<}_{mk}({\bf p}_3)
W^{<}_{l2}({\bf p}_4) \cr
&&-{W}^{<}_{1m}({\bf p}_1)W^{<}_{kl}({\bf p}_2)W^{>}_{mk}({\bf p}_3)
W^{>}_{l2}({\bf p}_4)] \}
\label{gamma12}
\end{eqnarray}
The above formula vanishes if $g_{11}=g_{12}=g_{22}$, i.e., the transverse
spin is conserved during collisions.
If the scattering length are not equal, collisions do not
conserve the transverse spin.

Assuming a small transverse spin component, i.e.  $|W_{12}|\ll W_{11},W_{22}$,
we expand $\Gamma_{n12}$ in $W_{12}$ to first order
\begin{eqnarray}
\Gamma_{n12}&=&\frac{2\pi}{\hbar}\int\frac{d{\bf p}_1}{(2\pi\hbar)^3}
\int\frac{d{\bf p}_2}{(2\pi\hbar)^3}\int\frac{d{\bf p}_3}{(2\pi\hbar)^3}
\int d{\bf p}_4 \cr
&&\times\delta(\tilde\varepsilon_{p_1}+\tilde\varepsilon_{p_2}
-\tilde\varepsilon_{p_3}-\tilde\varepsilon_{p_4})
\delta({\bf p}_1+{\bf p}_2-{\bf p}_3-{\bf p}_4) W_{12}({\bf p}_1)\cr
&&\times \Bigl[
g_{12}(g_{12}-g_{11})
\{[1+f_1({\bf p}_2)]f_1({\bf p}_3)f_2({\bf p}_4) \cr
&&-f_1({\bf p}_2)[1+f_1({\bf p}_3)][1+f_2({\bf p}_4)]\} \cr
&&+g_{22}(g_{22}-g_{12})\{
[1+f_2({\bf p}_2)]f_2({\bf p}_3)f_2({\bf p}_4) \cr
&&-f_2({\bf p}_2)[1+f_2({\bf p}_3)][1+f_2({\bf p}_4)]\}\cr
&&+g_{12}(g_{12}-g_{22})
\{[1+f_2({\bf p}_2)]f_2({\bf p}_3)f_1({\bf p}_4) \cr
&&-f_2({\bf p}_2)[1+f_2({\bf p}_3)][1+f_1({\bf p}_4)]\} \cr
&&+g_{11}(g_{11}-g_{12})\{
[1+f_1({\bf p}_2)]f_1({\bf p}_3)f_1({\bf p}_4) \cr
&&-f_1({\bf p}_2)[1+f_1({\bf p}_3)][1+f_1({\bf p}_4)]\}\Bigr],
\end{eqnarray}
where $f_i\equiv W_{ii}$.

To obtain a closed form in terms of the transverse spin component,
we assume the local equilibrium distribution given in Eqs.~(\ref{W_local})-(\ref{f_pm}).
Then $f_1,f_2$ and $W_{12}$ are written in terms of $f_+$ and $f_-$ as
\begin{equation}
f_1=\frac{1}{2}[(f_++f_-)+\cos\theta_n(f_+-f_-)],
\end{equation}
\begin{equation}
f_2=\frac{1}{2}[(f_++f_-)-\cos\theta_n(f_+-f_-)],
\end{equation}
\begin{equation}
W_{12}=\frac{1}{2}(f_+-f_-)\sin\theta_n e^{-i\phi_n}.
\end{equation}
With this form of $W_{12}$, we find
\begin{equation}
\left.\frac{\partial n_{12}}{\partial t}\right|_{\rm coll}=
-\frac{n_{12}}{\tau_2},
\end{equation}

In order to obtain the explicit form of the relaxation time $\tau_2$,
we consider the following two cases.
First, we assume that the spin direction is almost along the $z$
direction i.e., $\theta_n\approx 0$.
To first order in $\theta_n$, one has
\begin{equation}
f_1 \approx f_+, ~~ f_2\approx f_-, ~~
W_{12}\approx\frac{1}{2}(f_1-f_2)\theta_n e^{-i\phi_n}.
\end{equation}
The relaxation time $\tau_2$, which is a kind of ``$T2$" lifetime, is given by
\begin{eqnarray}
\frac{1}{\tau_2}&=&
\frac{2\pi}{(\tilde{n}_1-\tilde{n}_2)\hbar}\int\frac{d{\bf p}_1}{(2\pi\hbar)^3}
\int\frac{d{\bf p}_2}{(2\pi\hbar)^3}\int\frac{d{\bf p}_3}{(2\pi\hbar)^3}
\int d{\bf p}_4 \cr
&&\times \delta(\tilde\varepsilon_{p_1}+\tilde\varepsilon_{p_2}
-\tilde\varepsilon_{p_3}-\tilde\varepsilon_{p_4})
\delta({\bf p}_1+ {\bf p}_2-{\bf p}_3-{\bf p}_4) \cr
&&\times \Bigl[
(g_{11}-g_{12})^2\{[1+f_2({\bf p}_1)][1+f_1({\bf p}_2)]f_1({\bf p}_3)f_1({\bf p}_4)\cr
&&-f_2({\bf p}_1)f_1({\bf p}_2)[1+f_1({\bf p}_3)][1+f_1({\bf p}_4)]\}\cr
&&+(g_{22}-g_{12})^2\{[1+f_2({\bf p}_1)][1+f_2({\bf p}_2)]f_2({\bf p}_3)f_1({\bf p}_4)\cr
&&-f_2({\bf p}_1)f_2({\bf p}_2)[1+f_2({\bf p}_3)][1+f_1({\bf p}_4)]\}\Bigr].
\label{tau2_quantum}
\end{eqnarray}
If the spin is almost polarized along the longitudinal direction, i.e. 
$f_2\approx 0$, the above relaxation time reduces to
\begin{eqnarray}
\frac{1}{\tau_2}&=&
\frac{2\pi}{\tilde{n}_1\hbar}\int\frac{d{\bf p}_1}{(2\pi\hbar)^3}
\int\frac{d{\bf p}_2}{(2\pi\hbar)^3}\int\frac{d{\bf p}_3}{(2\pi\hbar)^3}
\int d{\bf p}_4 \cr
&&\times \delta(\tilde\varepsilon_{p_1}+\tilde\varepsilon_{p_2}
-\tilde\varepsilon_{p_3}-\tilde\varepsilon_{p_4})
\delta({\bf p}_1+ {\bf p}_2-{\bf p}_3-{\bf p}_4) \cr
&&\times 
(g_{11}-g_{12})^2[1+f_1({\bf p}_2)]f_1({\bf p}_3)f_1({\bf p}_4).
\label{tau2_1}
\end{eqnarray}

In the opposite limit of the small longitudinal polarization $f_1\approx
f_2$ and $\theta_n\approx \pi/2$, we write
\begin{equation}
\theta_n=\frac{\pi}{2}+\delta\theta_n,~~
\tilde\mu_+=\tilde\mu+\delta\tilde\mu,~~
\tilde\mu_-=\tilde\mu-\delta\tilde\mu.
\end{equation}
To first order in the small variables, one has
\begin{equation}
f_1\approx f_2\approx f,~~
W_{12}\approx \beta\delta\tilde\mu e^{-i\phi_n}f(1+f),
\end{equation}
with
\begin{equation}
f\equiv\frac{1}{e^{\beta(\tilde\varepsilon_p-\tilde\mu)}-1}
\equiv \frac{1}{z^{-1}e^{\beta(p^2/2m)}-1}. 
\end{equation}
We then obtain
\begin{eqnarray}
\frac{1}{\tau_2}&=&
\frac{g_{3/2}(z)}{g_{1/2}(z)}
\frac{4\pi}{\tilde{n}\hbar}\int\frac{d{\bf p}_1}{(2\pi\hbar)^3}
\int\frac{d{\bf p}_2}{(2\pi\hbar)^3}\int\frac{d{\bf p}_3}{(2\pi\hbar)^3}
\int d{\bf p}_4 \cr
&&\times \delta(\tilde\varepsilon_{p_1}+\tilde\varepsilon_{p_2}
-\tilde\varepsilon_{p_3}-\tilde\varepsilon_{p_4})
\delta({\bf p}_1+ {\bf p}_2-{\bf p}_3-{\bf p}_4) \cr
&&\times 
\left[(g_{11}-g_{12})^2+(g_{22}-g_{12})^2\right]\cr
&&\times[1+f({\bf p}_1)][1+f({\bf p}_2)]f({\bf p}_3)f({\bf p}_4),
\label{tau2_2}
\end{eqnarray}
where $\tilde n=\tilde n_1+\tilde n_2\approx (2/\lambda_T^3)g_{3/2}(z)$.
One can show that Eq.~(\ref{tau2_quantum}) smoothly goes over to Eq.~(\ref{tau2_2}),
if we take the limit $\tilde\mu_+\approx\tilde\mu_-\approx \tilde\mu$.

We now consider the simplest case of the high-temperature limit $T\gg T_c$.
In this limit, the local equilibrium distribution takes the Maxwell-Boltzmann
form
\begin{equation}
\uuline{W}^{\rm leq}({\bf r},{\bf p},t)\approx
\frac{1}{2}[\uuline{1}+\vec{M}_n({\bf r},t)\cdot\uuline{\vec{\sigma}}]
f_{\rm MB}({\bf r},{\bf p},t),
\end{equation}
with
\begin{equation}
f_{\rm MB}({\bf r},{\bf p},t)
\equiv\exp\left(-\beta\left[\frac{{\bf p}^2}{2m}
-\mu({\bf r},t)\right]\right).
\label{f_local}
\end{equation}
Assuming $f_{\rm MB}\ll 1$ and keeping the quadratic terms in Eq.~(\ref{gamma12}),
we obtain the following simple form of the relaxation time
\begin{eqnarray}
\frac{1}{\tau_2}&=&\frac{\pi}{\tilde{n}\hbar}
\int\frac{d{\bf p}_1}{(2\pi\hbar)^3}\int\frac{d{\bf p}_2}{(2\pi\hbar)^3}
\int\frac{d{\bf p}_3}{(2\pi\hbar)^3}d{\bf p}_4 \cr
&&\times \delta(\tilde\varepsilon_{p_1}+\tilde\varepsilon_{p_2}-\tilde\varepsilon_{p_3}
-\tilde\varepsilon_{p_4})
\delta({\bf p}_1+{\bf p}_2-{\bf p}_3-{\bf p}_4) \cr
&&\times\{[(g_{11}-g_{12})^2(1+M_n^z)
+(g_{22}-g_{12})^2(1-M_n^z)] \cr
&&\times f_{\rm MB}({\bf p}_3)f_{\rm MB}({\bf p}_4)\},
\label{tau2_classical}
\end{eqnarray}
where $\tilde n=(1/\lambda_T^3)e^{\beta\mu}$.
This expression for the relaxation time is formally identical to that of the
classical collision time, and thus one explicitly finds
\begin{equation}
\frac{1}{\tau_2}=\sqrt{2}n\sigma_{\rm eff}v_{\rm th},
\end{equation}
where $v_{\rm th}\equiv \sqrt{8k_{\rm B}T/\pi m}$ is the thermal
velocity and $\sigma_{\rm eff}$ is the effective collisional cross section
\begin{equation}
\sigma_{\rm eff}=2\pi[(a_{11}-a_{12})^2(1+M_n^z)+
(a_{22}-a_{12})^2(1-M_n^z)],
\end{equation}
rather than the usual $\sigma_{\rm coll}=8\pi a^2$.
One finds that this relaxation time is much longer than the mean collision
time $\tau_c=\sqrt{2}n\sigma_{\rm coll}v_{\rm th}$:
\begin{equation}
\tau_2/\tau_c \sim (a/\Delta a)^2,
\end{equation}
where $\Delta a$ is the difference in the scattering lengths.
For $^{87}$Rb, one finds $\tau_2/\tau_c\sim 1000$.
In the recent JILA experiment \cite{Lewandowski2002a}, this gives $\tau_2\sim
10$s.

We note that if $|g_{11}-g_{12}|=|g_{22}-g_{12}|$, which is the case of $^{87}$Rb,
$\tau_2$ in the classical limit is independent of the longitudinal polarization $M_n^z$.
However, in the quantum degenerate regime, we see from Eq.~(\ref{tau2_quantum})
that $\tau_2$ may strongly depend on $M_n^z$.
This effect might be observable in experiments near $T_c$.

We next briefly discuss the case of below $T_c$.
Since the $\uuline{I}{}_n$ collision integral has no explicit dependence on the condensate,
$\uuline{\Gamma}{}_n$ is the same as discussed above for $T>T_c$.
For $\Gamma_c$, integrating Eq.~(\ref{Ic_general}) over momentum, we find
\begin{equation} 
\uuline{\Gamma}{}_c=\uuline{\Gamma}{}_c^{(1)}+\uuline{\Gamma}{}_c^{(2)},
\end{equation}
where
\begin{equation}
\uuline{\Gamma}{}_c^{(1)}=\frac{1}{\hbar}\{\uuline{n}{}_c,\uuline{R}\}
=-\left.\frac{\partial\uuline{n}{}_c}{\partial t}\right|_{\rm coll},
\end{equation}
and
\begin{eqnarray}
\uuline{\Gamma}{}_{cij}^{(2)}&=&\frac{\pi}{\hbar}\int\frac{d{\bf p}_1}{(2\pi\hbar)^3}
\int\frac{d{\bf p}_2}{(2\pi\hbar)^3}\int d{\bf p}_3 \cr
&&\times\delta(\varepsilon_c+\tilde\varepsilon_{p_1}-\tilde\varepsilon_{p_2}
-\tilde\varepsilon_{p_3})\delta(m{\bf v}_c+{\bf p}_1-{\bf p}_2-{\bf p}_3) \cr
&&\times\sum_{mkl}g_{ml}(g_{kj}-g_{ik}) \cr
&&\times\bigl\{n_{ckl}\bigl[
W_{im}^{>}({\bf p}_1)W_{mj}^{<}({\bf p}_2)W_{lk}^<({\bf p}_3)
-W_{im}^<({\bf p}_1)W_{mj}^>({\bf p}_2)W_{lk}^>({\bf p}_3)\cr
&&+W_{im}^{>}({\bf p}_1)W_{mk}^{<}({\bf p}_2)W_{lj}^<({\bf p}_3)
-W_{im}^<({\bf p}_1)W_{mk}^>({\bf p}_2)W_{lj}^>({\bf p}_3)\bigr]\cr
&&+n_{clk}[W_{mj}^>({\bf p}_1)W_{kl}^<({\bf p}_2)W_{im}^<({\bf p}_3)
-W_{mj}^<({\bf p}_1)W_{kl}^>({\bf p}_2)W_{im}^<({\bf p}_3) \cr
&&+W_{mj}^>({\bf p}_1)W_{il}^<({\bf p}_2)W_{km}^<({\bf p}_3)
-W_{mj}^<({\bf p}_1)W_{il}^>({\bf p}_2)W_{km}^<({\bf p}_3) \bigr]\bigr\}.
\end{eqnarray}
One immediately finds that $\Gamma_{cii}^{(2)}=0$.
This means that collisions between the condensate and noncondensate
conserve the total density $n=n_1+n_2$ and the longitudinal spin 
$S^z=n_1-n_2$, but do not conserve the transverse component of the
total spin.
Further explicit calculations of $\uuline{\Gamma}{}_c$ in 
the unequal scattering length case is quite complicated, and thus we
do not make such analysis in the present paper.

\section{Conclusions}
\label{sec:conclusions}
In this paper, we derived a set of equations which describes the dynamics of a trapped
Bose-condensed gas with spin-1/2 internal degrees of freedom at finite temperatures.
These equations consist of a generalized Gross-Pitaevskii (GP) equation for the spinor
condensate order parameter $\underline{\Phi}({\bf r},t)$, as given by Eq.~(\ref{spinor_GP}),
and a semiclassical kinetic equation for the noncondensate distribution function
$\uuline{W}({\bf r},{\bf p},t)$ in a $2\times 2$ matrix form, as given by Eq.~(\ref{KE}).
These coupled equations are the spin-1/2 generalization of the
kinetic equations obtained by ZNG for a single component gas~\cite{Zaremba1999a}, which has been
shown to accurately describe several different experiments at finite temperatures~\cite{Jackson2002a}. 
Both of these theories treat the thermal excitations semiclassically in the Hartree-Fock approximation.
Our kinetic theory is also the natural extension to the Bose-condensed regime of the earlier theories
developed by Jeon and Mullin~\cite{Jeon1988a,Jeon1989a,Mullin1992a} and Ruckenstein and
L\' evy~\cite{Ruckenstein1989a} to describe spin waves in dilute quantum degenerate gases.

As a further development of our theory, from Eq.~(\ref{spinor_GP}) and Eq.~(\ref{KE}) we derived
a closed set of spin hydrodynamic equations for the two spin components in the absence of any
overall mass currents, i.e. ${\bf v}_c={\bf v}_n=0$, as summarized in 
Eqs.~(\ref{spinhydro1})-(\ref{spinhydro4}). 
These equations are based on the assumption that collisions between noncondensate atoms 
(described by the $\uuline{I}{}_n$ collision integral) are sufficiently rapid to produce local equilibrium.
Deviation from local equilibrium gives rise to the usual spin-diffusion relaxation in the
noncondensate spin current, which now involves an additional contribution from the collisions
of noncondensate atoms with atoms in the condensate (described by the $\uuline{I}{}_c$ collision integral).
Interestingly, the $\uuline{I}{}_c$ collision integral also gives rise the mutual relaxation
between the condensate and noncondensate spins, which suggests that the condensate and noncondensate
spins have a tendency to become aligned. This is a new prediction of our theory. In addition to the 
collisional coupling between the spins of the condensate and noncondensate, the exchange mean
field gives rise to a mutual torque coupling the two spins. This mean field effect was first pointed out 
by Oktel and Levitov~\cite{Oktel2002c}.

There are a wealth of different applications of the present kinetic theory to finite temperature
dynamics of a spin-1/2 Bose-condensed gas. Perhaps the most obvious application is
to study the collective spin-wave dynamics below $T_c$, which was never treated in the
earlier literature on spin waves in spin polarized gases
\cite{Jeon1988a,Jeon1989a,Mullin1992a,Ruckenstein1989a}. 
From earlier studies of spin waves above $T_c$, we know that the exchange mean field has a
strong effect on the collective dynamics of the transverse spin in the thermal gas due to the
molecular field contribution to $\vec \Omega_n = \vec \Omega_n' + g_{12}\vec S_n/\hbar$.
Below $T_c$, there is an additional contribution to the molecular field experienced by the
noncondensate due to the condensate $\vec \Omega_n = \vec \Omega_n' + g_{12}(\vec S_n+ \vec S_c)/\hbar$.
The molecular field experienced by the condensate has a reciprocal coupling to the noncondensate
$\vec \Omega_c = \vec \Omega_c' + g_{12}\vec S_n/\hbar$, but notice that the condensate
by itself does not experience the exchange mean field: interactions do not
lead to collective spin waves in the condensate.
Instead, spin waves in the condensate will result
from the kinetic energy cost to twist up the condensate, an effect that is encapsulated by the
so-called ``quantum torque" form of the condensate spin current 
${\bf \vec J}_c=-(\hbar/2 n_c)\vec{S}_c\times{\boldsymbol\nabla}\vec{S}_c$.
Together with the spin relaxation effects due to collisions,
these various effects should lead to interesting new spin-wave dynamics below $T_c$
\cite{McGuirk2003a}.

Another important application, motivated by recent experimental observations~\cite{Lewandowski2003a}
is to study the effect that the dephasing of the spins in the thermal cloud has on the spinor condensate
after a $\pi/2$ pulse is applied. 
In the presence of spatial inhomogeneities in the external field
$\vec{\Omega}_n'({\bf r})$, the spin of the
noncondensate can decay to zero $\vec S_n({\bf r})\rightarrow 0$ due to the dephasing of individual spins 
of atoms as they travel with different trajectories in the trap. In contrast, all the atoms in the condensate are
in the same spinor wavefunction and so it is always spin polarized in our theory. As the spins of
the thermal gas dephase, the gas evolves from a spin coherent sample to a $50-50$ spin mixture,
which has a condensation temperature approximately $20\%$ lower than that of the spin polarized gas
\cite{Williams2003a}. 
At fixed temperature, we then expect the condensate fraction to decrease as the noncondensate spin
decoheres.
This ``condensate melting" \cite{Lewandowski2003a} is a finite temperature effect that requires a 
kinetic theory description that treats the scattering of atoms between the condensate and noncondensate,
and so is an ideal application of the present theory.
A related question arising from this study concerns the validity of the present kinetic theory for
describing the transverse spin dynamics of the
condensate when it is subjected to such strong phase fluctuations.
Indeed, two recent studies of the ground state of a spin-1/2 Bose gas suggest that in certain limits,
the condensate may become fragmented~\cite{Kuklov2002b,Ashhab2003a}, 
which is beyond our treatment of the condensate in the present kinetic theory.

A final proposed application of the theory is to treat the dynamics of topological states,
or spin textures, of the condensate at finite temperatures.
Several recent theoretical studies have explored spin textures at zero
temperature~\cite{Garcia-Ripoll2000a,Khawaja2001a}. 
We note that the two-component vortex state proposed~\cite{Williams1999c} and 
observed~\cite{Matthews1999b} at JILA is also a spin texture. 
At finite temperatures, based on our kinetic theory, one expects interesting effects due to the
mean field and collisional coupling between the spins of the condensate and thermal gas. 
In particular, the same kind of mutual friction type effects arising in a single component
Bose-condensed gas when the condensate and noncondensate velocities differ~\cite{Williams2002b} 
will also occur in a spin-1/2 gas. 
The generalization of this effect on spin dynamics
should follow from our theory. 
With the kinetic theory we have presented, one should be able to explore the role of the thermal cloud
in the creation and stability of topological excitations, as was done in the study of vortex formation
in a single component  gas~\cite{Williams2002b}.

\section{Acknowledgements}
We would like to thank the following people for useful discussions about this work during the 
last year and a half:
G. Brennen, C. W. Clark, E. A. Cornell, Z. Dutton, J. N. Fuchs, A. Griffin, D. M. Harber, B. Jackson,
F. Lalo\"e, H. J. Lewandowski, J. M. McGuirk, N. Nygaard, B. Statt, and E. Zaremba.

\begin{appendix}
\section{Two-level systems in dilute atomic gases}
\label{two-level}
With a bosonic atom, the two-levels making up the spin-1/2 system will always be a particular 
subspace of a larger space of internal states.
In the alkali atoms, this larger space is the set of hyperfine states, illustrated in Fig.~1.
In order to be a viable candidate, the system should meet a few basic requirements:
the two-levels should have very low loss rates and long coherence times;
it should be straight-forward to couple the states with a radiation field;
they must also experience the same confinement while still allowing for spin-dependent forces that
can excite collective modes of the spin.

There will always be collisions that take the system out of the two-level subspace, which we will consider as loss processes.
Typically, the most severe source of loss comes from spin-exchange collisions
\cite{Julienne1997a,Kokkelmans1997a,Burke1997a} (dipolar and three-body losses, which should be
examined as well, are typically much smaller).
The best way to minimize this loss process is to choose two states for which the selection rules or energetics do not allow the colliding atoms to change their internal states to ones outside the subspace.
In these spin changing collisions, the total molecular spin $F_{\rm{tot}}$ and spin projection $M_{\rm{tot}}$ of the colliding pair are conserved.
For example, for colliding atoms in the states $(1,1)$ and $(1,-1)$, the total spin can add up to the values $F_{\rm{tot}}=0,1,2$ and the spin projection is $M_{\rm{tot}}=0$.
Although the selection rules allow for scattering into states in the upper $F=2$ hyperfine manifold, these collisions are disallowed energetically at ultralow temperatures.
However, the process $(1,1)+(1,-1) \rightarrow (1,0)+(1,0)$ is allowed, making these two states a poor spin-1/2 candidate in general.
Based on similar arguments, one can show that spin exchange losses can be eliminated by using either of the two pairs of stretched states: $(1,-1)$ and $(2,-2)$ or $(1,1)$ and $(2,2)$, which can be coupled using a single photon microwave coupling.
For ${}^{87}$Rb, any pair of hyperfine states may be considered since
the spin exchange loss rates are anomalously low.

In order to confine the two states within the same magnetic trap, they must have the same magnetic
moment.
In this regard, the stretched states $(1,-1)$ and $(2,-2)$ or $(1,1)$ and $(2,2)$ would be poor
choices in a magnetic trap.
It seems that the states $(1,-1)$ and $(2,1)$ of ${}^{87}$Rb may be the only viable spin-1/2 system in a magnetic trap.
In order to use the stretched states, an optical trap must be used, which exerts the same force on
all of the hyperfine states.
In general there are various possible schemes for generating spin-dependent forces; a detailed
discussion is beyond the scope of this paper.

Based on these considerations, we have narrowed the possible two-state systems to either pair of
stretched states $(1,-1)$ and $(2,-2)$ or $(1,1)$ and $(2,2)$ for the alkali atoms
${}^{7}$Li, ${}^{23}$Na, ${}^{87}$Rb in an optical trap, or the states $(1,-1)$ and $(2,1)$ of
${}^{87}$Rb in a magnetic trap \cite{Matthews1999c,Williams1999d}.
These are illustrated in Fig.~1.
We note that at present, only the system in Fig.~1a has been explored in any detail in actual
experiments, but the spin-1/2 systems in Fig.~1b would allow for a wider range of parameters to
be explored (such as the effects of largely different scattering lengths in ${}^{23}$Na or
having an attractive interaction in ${}^{7}$Li).

\section{Derivation of collision integrals}
\label{sec:collisions}

In this Appendix, we give detailed derivation of the collision integral $\uuline{I}{}_n$
and $\uuline{I}{}_c$ given in Eqs.(\ref{Ic_symmetric}) and (\ref{In_symmetric}).
We closely follow the approach of KD \cite{Kirkpatrick1985c} and ZNG \cite{Zaremba1999a}.
As usual, we assume that the macroscopic variables are slowly
varying in space and time compared to the spatial and temporal scale of
a collision event. We thus approximate the macroscopic variables near $({\bf r},t)$ as
\begin{eqnarray}
&&\tilde n_{ij}({\bf r}',t')\approx\tilde n_{ij}({\bf r},t),~~
U_n({\bf r}',t')\approx U_n({\bf r},t), \cr
&&n_c({\bf r'},t')\approx n_c({\bf r},t),~~
U_c({\bf r}',t')\approx U_c({\bf r},t)
\label{local_approx}
\end{eqnarray}
For the condensate order parameter, we expand the three phases
$\alpha_c,\phi_c,\theta_c$ to first order in space and time:
\begin{equation}
\alpha_c({\bf r}',t')\approx \alpha_c({\bf r},t)
+\frac{\partial\alpha_c}{\partial t}(t'-t)
+{\boldsymbol\nabla}\alpha_c\cdot({\bf r}'-{\bf r})
\equiv \alpha_c({\bf r},t)+\delta\alpha_c,
\end{equation}
\begin{equation}
\phi_c({\bf r}',t')\approx \phi_c({\bf r},t)
+\frac{\partial\phi_c}{\partial t}(t'-t)
+{\boldsymbol\nabla}\phi_c\cdot({\bf r}'-{\bf r})
\equiv \phi_c({\bf r},t)+\delta\phi_c,
\end{equation}
\begin{equation}
\theta_c({\bf r}',t')\approx \theta_c({\bf r},t)
+\frac{\partial\theta_c}{\partial t}(t'-t)
+{\boldsymbol\nabla}\theta_c\cdot({\bf r}'-{\bf r})
\equiv \theta_c({\bf r},t)+\delta\theta_c,
\end{equation}
From Eq.~(\ref{cond_spinor}) we then obtain
\begin{eqnarray}
\underline{\Phi}({\bf r}',t')&\approx& 
e^{i\delta\alpha_c}
\left[\underline{1}
-i\frac{1}{2}\delta\phi_c\uuline{\sigma}^z
-i\frac{1}{2}\delta\theta_c(\cos\phi_c\uuline{\sigma}^y
-\sin\phi_c\uuline{\sigma}^x)\right]\underline{\Phi}\cr
&\simeq&\exp \left\{i\left[\delta\alpha_c\uuline{1}
-\frac{1}{2}\delta\phi_c\uuline{\sigma}^z-\frac{1}{2}\delta\theta_c
(\cos\phi_c\uuline{\sigma}^y-\sin\phi_c\uuline{\sigma}^x )\right]\right\}
\underline{\Phi} \cr
&=&\exp \left\{i\left[
\delta\alpha_c
\uuline{1}
-\frac{1}{2}\delta\phi_c\vec{M}_c\cdot\vec{\uuline{\sigma}}\cos\theta_c
-\frac{1}{2}\vec{\uuline{\sigma}}\cdot(\vec{M}_c\times
\delta\vec{M}_c)\right]\right\}\underline{\Phi},
\label{Phi_expand}
\end{eqnarray}
where
\begin{equation}
\delta\vec{M}_c\equiv\frac{\partial\vec{M}_c}{\partial t}(t'-t)
+{\boldsymbol\nabla}\vec{M}_c\cdot({\bf r}'-{\bf r}).
\end{equation}
One can also write Eq.~(\ref{Phi_expand}) as
\begin{equation}
\underline{\Phi}({\bf r}',t')\approx
\exp\left\{ 
\frac{i}{\hbar}\left[
\uuline{{\bf p}}{}_c \cdot({\bf r}'-{\bf r})
-\uuline{\varepsilon}{}_c(t'-t)\right]\right\}
\underline{\Phi}({\bf r},t),
\end{equation}
where
\begin{equation}
\uuline{{\bf p}}{}_c\equiv \hbar\left\{
{\boldsymbol\nabla}\alpha_c\uuline{1}
-\frac{1}{2}\vec{\uuline{\sigma}}\cdot
\left[\vec{M}_c{\boldsymbol\nabla}\phi_c\cos\theta_c
+\vec{M}_c\times{\boldsymbol\nabla}\vec{M}_c\right]\right\},
\end{equation}
\begin{equation}
\uuline{\varepsilon}{}_c\equiv -\hbar\left\{
\frac{\partial\alpha_c}{\partial t}\uuline{1}
-\frac{1}{2}\vec{\uuline{\sigma}}\cdot
\left[\vec{M}_c\frac{\partial\phi_c}{\partial t}\cos\theta_c
+\vec{M}_c\times\frac{\partial \vec{M}_c}{\partial t}\right]\right\},
\end{equation}
The spatial gradients of the phases are related to the condensate
velocity ${\bf v}_c$ through Eq.~(\ref{vc_def}).
Inserting the spinor form of the condensate wavefunction into the generalized
GP equation Eq.~(\ref{spinor_GP}), we also find
\begin{equation}
\hbar\left(\frac{\partial\alpha_c}{\partial t}-\frac{1}{2}\cos\theta_c
\frac{\partial\phi_c}{\partial t}\right)=-\varepsilon_c,
\end{equation}
where the condensate energy $\varepsilon_c$ is defined by (we neglect the anomalous
correlation $\tilde m_{ij}$)
\begin{eqnarray}
\varepsilon_c({\bf r},t)&\equiv&-\frac{\hbar^2\nabla^2\sqrt{n_c({\bf r},t)}}
{2m\sqrt{n_c({\bf r},t)}}+U_c({\bf r},t) \cr
&&+\frac{\hbar}{2}\vec{\Omega}_c
({\bf r},t)\cdot\vec{M}_c({\bf r},t)
+\frac{\hbar^2}{8m}[\boldsymbol\nabla\vec{M}_c({\bf r},t)]^2 \cr
&&+\frac{1}{n_c({\bf r},t)}\sum_{ij}g_{ij}{\rm Re}
\Bigl[
\Phi^*_i({\bf r},t)\langle\tilde\psi^{\dagger}_j({\bf r},t)\tilde\psi_j({\bf r},t)
\tilde\psi_i({\bf r},t)\rangle\Bigr] \cr
&&+\frac{mv_c^2({\bf r},t)}{2}\cr
& \equiv& \mu_c({\bf r},t)+\frac{mv_c^2({\bf r},t)}{2},
\label{eq_muc}
\end{eqnarray}
where $\mu_c({\bf r},t)$ is the local condensate chemical potential.

It is now convenient to introduce the Fourier transform
\begin{equation}
\tilde\psi_i({\bf r},t_0)=\frac{1}{\sqrt{V}}\sum_{{\bf p}}a_{i{\bf p}}
e^{i{\bf p}\cdot{\bf r}/\hbar},~~
\tilde\psi^{\dagger}_i({\bf r},t_0)=\frac{1}{\sqrt{V}}\sum_{{\bf p}}
a^{\dagger}_{i{\bf p}}e^{-i{\bf p}\cdot{\bf r}/\hbar},
\end{equation}
where $V$ is the volume of the system.
For convenience on calculations, we work with the finite volume $V$
and thus treat discrete momentum variable ${\bf p}$.
After deriving collision terms, we take the limit $V\to\infty$
to replace the summation over ${\bf p}$ with the momentum integral.
With the local approximation in Eq.~(\ref{local_approx}), 
one can write the mean-field Hamiltonian $\hat H_0(t')$ as
\begin{equation}
\hat H_0(t')\approx\sum_{i,j}\sum_{\bf p}\left\{\left[\frac{p^2}{2m}+U_n({\bf r},t)\right]
\delta_{ij}+\frac{\hbar}{2}\langle i|\vec{\Omega}_n({\bf r},t)\cdot\vec{\uuline{\sigma}}|j
\rangle \right\}
a^{\dagger}_{i{\bf p}}a_{j{\bf p}}.
\label{H0_local}
\end{equation}
With this approximate form of $\hat H_0(t')$, the free evolution operator
becomes
\begin{equation}
\hat\EuScript{U}_0(t,t') 
\approx \exp\left[-\frac{i}{\hbar}\hat H_0(t-t')\right].
\end{equation}
In performing the perturbation calculation using Eq.~(\ref{2nd-order}),
one needs to evaluate $\hat\EuScript{U}_0^{\dagger}(t,t')
a_{i{\bf p}}\hat\EuScript{U}_0(t,t')$.
It can be formally written as
\begin{eqnarray}
&&\hat\EuScript{U}_0^{\dagger}(t,t') 
a_{i{\bf p}}\hat\EuScript{U}_0(t,t') \cr
&&=
\sum_j
\langle i| \exp\left\{-\frac{i}{\hbar} \uuline{H}{}_n({\bf r},{\bf p},t)(t-t')\right\}
|j\rangle
a_{j{\bf p}}.
\label{at_formal}
\end{eqnarray}
In principle, one can diagonalize Eq.~(\ref{H0_local}) to write down an explicit
solution for Eq.~(\ref{at_formal}).
However, doing so would be quite complicated.
Following Jeon and Mullin~\cite{Jeon1988a}, 
we use the simple approximation of
introducing the local HF excitation energy $\tilde\varepsilon_p$,
\begin{equation}
\hat\EuScript{U}_0^{\dagger}(t,t')a_{i{\bf p}}\hat\EuScript{U}_0(t,t')\approx 
\exp\left[-\frac{i}{\hbar}
\tilde\varepsilon_p(t-t')\right]a_{i{\bf p}},
\label{at_approx}
\end{equation}
where
\begin{equation}
\tilde\varepsilon_p({\bf r},t)\equiv \frac{p^2}{2m}+U_n({\bf r},t)
+\frac{\hbar}{2}\vec{\Omega}_n({\bf r},t)\cdot\vec{M}_n({\bf r},t).
\label{HF_excitation}
\end{equation}
The above approximation corresponds to replacing $\vec{\uuline{\sigma}}$ with
$\vec{M}_n\uuline{1}$.
We also use the same level of approximation for  the condensate order parameter:
\begin{equation}
\underline{\Phi}({\bf r}',t')\approx
\exp\left\{i\left[-\frac{\varepsilon_c}{\hbar}(t'-t)+
\frac{1}{\hbar}{\bf p}_c\cdot({\bf r}'-{\bf r})\right]\right\}\underline{\Phi},
\label{Phirt_approx}
\end{equation}
where the local condensate momentum is defined by
${\bf p}_c({\bf r},t)\equiv m{\bf v}_c({\bf r},t)$.
We can also express $\hat H'(t')$ in terms of the Fourier transform:
\begin{eqnarray}
\hat H'_1(t')&\approx&-V^{1/2}\sum_{ij}\sum_p \delta_{{\bf p},{\bf p}_c}
\cr 
&&\times g_{ij} \left[e^{i{\bf p}_c\cdot{\bf r}/\hbar}e^{-i\varepsilon_c(t-t')/\hbar}
(\Phi^{*}_i\tilde n_{jj}+\Phi_j^{*}\tilde n_{ji})a_{i{\bf p}}+{\rm H.c.}\right],
 \\
\hat H'_2(t')&\approx& \sum_{ij}\sum_{{\bf p}_1,{\bf p}_2}
\delta_{{\bf p}_1+{\bf p}_2,2{\bf p}_c} \cr
&&\times \frac{g_{ij}}{2}
\left[e^{i2{\bf p}_c\cdot{\bf r}/\hbar}e^{-i2\varepsilon_c(t-t')/\hbar}
\Phi^{*}_{i}\Phi^*_ja_{j{\bf p}_1}a_{i{\bf p}_2}+{\rm H.c.}
\right], \\
\hat H'_3(t')&\approx& \frac{1}{V^{1/2}}\sum_{ij}\sum_{p_1,p_2,p_3}
\delta_{{\bf p}_c+{\bf p}_1,{\bf p}_2+{\bf p}_3}\cr
&&\times g_{ij}\left[e^{i{\bf p}_c\cdot{\bf r}/\hbar}e^{-i\varepsilon_c(t-t')/\hbar}
\Phi_i^*a_{j{\bf p}_1}^{\dagger}a_{j{\bf p}_2}a_{i{\bf p}_3}+
{\rm H.c.}\right], \\
\hat H'_4(t')&\approx& \frac{1}{2V}\sum_{ij}\sum_{{\bf p}_1,{\bf p}_2,
{\bf p}_3,{\bf p}_4}\delta_{{\bf p}_1+{\bf p}_2,{\bf p}_3+{\bf p}_4}
g_{ij}a^{\dagger}_{i{\bf p}_1}a^{\dagger}_{j{\bf p}_2}a_{j{\bf p}_3}a_{i{\bf p}_4}
\cr
&&-\sum_{ij}\sum_{{\bf p}}g_{ij}(\tilde n_{ji}a^{\dagger}_{j{\bf p}}a_{i{\bf p}}
+\tilde n_{jj}a^{\dagger}_{i{\bf p}}a_{i{\bf p}}).
\end{eqnarray}

As usual, we assume that there is no initial correlation so that we can
use the Wick's theorem to factorize terms such as
\begin{eqnarray}
&&\langle a_{i{\bf p}_1}^{\dagger}a_{j{\bf p}_2}^{\dagger}
a_{k{\bf p}_3}a_{l{\bf p}_4}\rangle_{t_0} \cr
&&=\langle a_{i{\bf p}_1}^{\dagger}a_{k{\bf p}_3}\rangle_{t_0}
\langle a_{j{\bf p}_2}^{\dagger} a_{l{\bf p}_4}\rangle_{t_0}+
\langle a_{i{\bf p}_1}^{\dagger}a_{l{\bf p}_4}\rangle_{t_0}
\langle a_{j{\bf p}_2}^{\dagger} a_{k{\bf p}_3}\rangle_{t_0}.
\end{eqnarray}
As we noted before, the anomalous correlations are neglected.
Within the present approximation of slow variation of the macroscopic
quantities in space and time, we can use
\begin{equation}
\langle a^{\dagger}_{j{\bf p}_1}a_{i{\bf p}_2}\rangle_{t_0} \approx
\delta_{{\bf p}_1,{\bf p}_2}W_{ij}({\bf r},{\bf p},t),
\end{equation}
in the calculation of the collision integral.

We first consider the three-field correlation function that appears in the
generalized GP equation:
\begin{equation}
\langle\tilde\psi^{\dagger}_i({\bf r},t)\tilde\psi_j({\bf r},t)
\tilde\psi_k({\bf r},t)\rangle
=\frac{1}{V^{3/2}}\sum_{{\bf p}_1,{\bf p}_2,{\bf p}_3}
\langle a^{\dagger}_{i{\bf p}_1}a_{j{\bf p}_2}a_{k{\bf p}_3}\rangle_t
e^{-i({\bf p}_1-{\bf p}_2-{\bf p}_3)\cdot{\bf r}/\hbar}.
\end{equation}
Clearly the first term in Eq.~(58) makes no contribution to the three-field
correlation function.
We then find
\begin{eqnarray}
&&\langle a^{\dagger}_{i{\bf p}_1}a_{j{\bf p}_2}a_{k{\bf p}_3}\rangle_t \cr
&&=-\frac{i}{\hbar}\int_{t_0}^t dt' 
\langle \hat\EuScript{U}_0^{\dagger}(t',t_0)
[\hat\EuScript{U}_0^{\dagger}(t,t')a^{\dagger}_{i{\bf p}_1}a_{j{\bf p}_2}a_{k{\bf p}_2}
\hat\EuScript{U}_0(t,t'),\hat H'(t')]\hat\EuScript{U}_0(t',t_0)\rangle_{t_0} \cr
&&=-\frac{i}{\hbar}\int_{t_0}^tdt' e^{i(\tilde\varepsilon_{p_1}
-\tilde\varepsilon_{p_2}-\tilde\varepsilon_{p_3})(t-t')/\hbar} \cr
&&~~~~~~~~\times \langle [a^{\dagger}_{i{\bf p}_1}a_{j{\bf p}_2}a_{k{\bf p}_2},
\hat H'_1(t')+H'_3(t')]\rangle_{t_0}.
\end{eqnarray}
The time integral is performed by setting $t_0\to-\infty$ introducing the
convergence factor $e^{-\delta(t-t')}$, which yields
\begin{eqnarray}
&&\frac{1}{\hbar}\int_{-\infty}^t dt' e^{i(\varepsilon_c+\tilde\varepsilon_{p_1}-
\tilde\varepsilon_{p_2}-\tilde\varepsilon_{p_3})(t-t')/\hbar} \cr
&&=\pi\delta(\varepsilon_c+\tilde\varepsilon_{p_1}-\tilde\varepsilon_{p_2}
-\tilde\varepsilon_{p_3})+i{\cal P}\left(\frac{1}{\varepsilon_c+\tilde\varepsilon_{p_1}
-\tilde\varepsilon_{p_2}-\tilde\varepsilon_{p_3}}\right)
\end{eqnarray}
Using the Wick's theorem, we finally obtain
\begin{eqnarray}
&&\langle a^{\dagger}_{i{\bf p}_1}a_{j{\bf p}_2}a_{k{\bf p}_3}\rangle_t \cr
&&=
-i\left[\pi\delta(\varepsilon_c+\tilde\varepsilon_{p_1}-\tilde\varepsilon_{p_2}
-\tilde\varepsilon_{p_3})+i{\cal P}\left(\frac{1}{\varepsilon_c+\tilde\varepsilon_{p_1}
-\tilde\varepsilon_{p_2}-\tilde\varepsilon_{p_3}}\right)\right] \cr
&&e^{-i{\bf p}_c\cdot{\bf r}/\hbar}\delta_{{\bf p}_c+{\bf p}_1,{\bf p}_2+{\bf p}_3}
\cr && \times \frac{1}{\sqrt{V}}\sum_{lm}g_{lm}\Phi_l 
\bigl\{
W_{mi}({\bf p}_1)[\delta_{jm}+W_{jm}({\bf p}_2)][\delta_{kl}+W_{kl}({\bf p}_3)]\cr
&&+ W_{mi}({\bf p}_1)[\delta_{jl}+W_{jl}({\bf p}_2)][\delta_{km}+W_{km}({\bf p}_3)] \cr
&&-[\delta_{mi}+W_{mi}({\bf p}_1)]W_{jm}({\bf p}_2)W_{kl}({\bf p}_3) \cr
&&-[\delta_{mi}+W_{mi}({\bf p}_1)]W_{jl}({\bf p}_2)W_{km}({\bf p}_3) \bigr\},
\label{3field_p}
\end{eqnarray}
where we omitted the arguments ${\bf r}$ and $t$ for simplicity.
We thus obtain the three-field correlation function
\begin{eqnarray}
&&\langle \tilde\psi^{\dagger}_i\tilde\psi_j\tilde\psi_k\rangle \cr
&&=
-i\frac{1}{V^2}\sum_{{\bf p}_1,{\bf p}_2,{\bf p}_3}
\left[\pi\delta(\varepsilon_c+\tilde\varepsilon_{p_1}-\tilde\varepsilon_{p_2}
-\tilde\varepsilon_{p_3})+i{\cal P}\left(\frac{1}{\varepsilon_c+\tilde\varepsilon_{p_1}
-\tilde\varepsilon_{p_2}-\tilde\varepsilon_{p_3}}\right)\right] \cr
&&\times \delta_{{\bf p}_c+{\bf p}_1,{\bf p}_2+{\bf p}_3}\sum_{lm}g_{lm}\Phi_l 
\bigl\{
W_{mi}({\bf p}_1)[\delta_{jm}+W_{jm}({\bf p}_2)][\delta_{kl}+W_{kl}({\bf p}_3)]\cr
&&+ W_{mi}({\bf p}_1)[\delta_{jl}+W_{jl}({\bf p}_2)][\delta_{km}+W_{km}({\bf p}_3)] \cr
&&-[\delta_{mi}+W_{mi}({\bf p}_1)]W_{jm}({\bf p}_2)W_{kl}({\bf p}_3) \cr
&&-[\delta_{mi}+W_{mi}({\bf p}_1)]W_{jl}({\bf p}_2)W_{km}({\bf p}_3) \bigr\},
\label{3field_r}
\end{eqnarray}
Using this result, one obtains the explicit formula for the three-field contribution
in the GP equation as
\begin{eqnarray}
&&\sum_j g_{ij}\langle \tilde\psi^{\dagger}_j\tilde\psi_j\tilde\psi_i\rangle \cr
&&=-i\frac{1}{V^2}\sum_{{\bf p}_1,{\bf p}_2,{\bf p}_3}
\left[\pi\delta(\varepsilon_c+\tilde\varepsilon_{p_1}-\tilde\varepsilon_{p_2}
-\tilde\varepsilon_{p_3})+i{\cal P}\left(\frac{1}{\varepsilon_c+\tilde\varepsilon_{p_1}
-\tilde\varepsilon_{p_2}-\tilde\varepsilon_{p_3}}\right)\right] \cr
&&\times \delta_{{\bf p}_c+{\bf p}_1,{\bf p}_2+{\bf p}_3}
[A_{ij}^{<}({\bf p}_1;{\bf p}_2,{\bf p}_3)-A_{ij}^{>}({\bf p}_1;{\bf p}_2,{\bf p}_3)]\Phi_j \cr
&&\equiv \sum_j R_{ij}\Phi_j,
\label{3field_GP}
\end{eqnarray}
where we have defined the matrices
\begin{eqnarray}
A_{ij}^{\stackrel{<}{>}}({\bf p}_1;{\bf p}_2,{\bf p}_3)&\equiv&\sum_{kl}
g_{kj}g_{il}\{W^{\stackrel{>}{<}}_{kl}({\bf p}_1)
W^{\stackrel{<}{>}}_{lk}({\bf p}_2)W^{\stackrel{<}{>}}_{ij}({\bf p}_3)\cr
&&~~+W^{\stackrel{>}{<}}_{kl}({\bf p}_1)W^{\stackrel{<}{>}}_{ik}({\bf p}_2)
W^{\stackrel{<}{>}}_{lj}({\bf p}_3)\},
\label{A<}
\end{eqnarray}
with
\begin{equation}
\uuline{W}^<({\bf p})\equiv\uuline{W}({\bf p}),~
\uuline{W}^>({\bf p})\equiv\uuline{1}+\uuline{W}({\bf p}).
\end{equation}
This notation using $>$ and $<$ follows the Kadanoff-Baym formalism
(actually, these matrices are directly related to the second-order
self-energy in the Kadanoff-Baym formalism) 
\cite{Kadanoff1962a,Imamovic-Tomasovic2001a,Jeon1988a}.
We note that the matrix $A^{\stackrel{<}{>}}_{ij}$ has the property
\begin{equation}
A^{\stackrel{<}{>}}_{ij}({\bf p}_1;{\bf p}_2,{\bf p}_3)
=[A^{\stackrel{<}{>}}_{ji}({\bf p}_1;{\bf p}_3,{\bf p}_2)]^*
\end{equation}
Eq.~(\ref{3field_GP}) defines the matrix $\uuline{R}$ that appears in the generalized GP equation
Eq.~(\ref{spinor_GP}).

The similar techniques and approximations are used to evaluate the
collision terms in the kinetic equation.
We first consider the contribution from $\hat H'_3$:
\begin{eqnarray}
&&-\frac{i}{\hbar}{\rm tr}\tilde\rho(t,t_0)
[\hat{W}_{ij} ({\bf r},{\bf p},t),\hat H_3'(t)] \cr
&&\simeq -i\frac{1}{\hbar}\frac{1}{\sqrt{V}}\sum_{\bf q}
\sum_{{\bf p}_1,{\bf p}_2,{\bf p}_3}\delta_{{\bf p}_c+{\bf p}_1,
{\bf p}_2+{\bf p}_3} \cr
&&\times \bigl\{e^{i{\bf p}_c\cdot{\bf r}/\hbar}
\sum_k[g_{ik}\delta_{{\bf p},{\bf p}_1-{\bf q}/2}\Phi^{*}_k
\langle a^{\dagger}_{j{\bf p}-{\bf q}/2}a_{i{\bf p}_2}a_{k{\bf p}_3}\rangle_t\cr
&&-g_{kj}\delta_{{\bf p},{\bf p}_2+{\bf q}/2}\Phi^{*}_k
\langle a^{\dagger}_{j{\bf p}}a_{i{\bf p}+{\bf q}/2}a_{k{\bf p}_3}\rangle_t 
-g_{kj}\delta_{{\bf p},{\bf p}_3+{\bf q}/2}\Phi^{*}_j
\langle a^{\dagger}_{k{\bf p}}a_{k{\bf p}_2}a_{i{\bf p}+{\bf q}/2}\rangle_t ]\cr
&&-e^{i{\bf p}_c\cdot{\bf r}/\hbar}\sum_k[\delta_{{\bf p}_1,{\bf p}-{\bf q}/2}
g_{kj}\Phi_k\langle a^{\dagger}_{k{\bf p}_3}a^{\dagger}_{j{\bf p}_2}
a_{i{\bf p}+{\bf q}/2}\rangle_t \cr
&&-\delta_{{\bf p}_2,{\bf p}+{\bf q}/2}g_{ik}\Phi_k
\langle a^{\dagger}_{j{\bf p}-{\bf q}/2}a^{\dagger}_{k{\bf p}_3}a_{i{\bf p}_1}\rangle_t
\cr
&&-\delta_{{\bf p}_3,{\bf p}+{\bf q}/2}g_{ik}\Phi_i
\langle a^{\dagger}_{j{\bf p}-{\bf q}/2}a^{\dagger}_{k{\bf p}_2}a_{k{\bf p}_1}\rangle_t
]\bigr\} \cr
&&\simeq-i\frac{1}{\hbar}\frac{1}{\sqrt{V}} \sum_{{\bf p}_1,{\bf p}_2,{\bf p}_3}
\delta_{{\bf p}_c+{\bf p}_1,{\bf p}_2+{\bf p}_3}\sum_k \cr
&&\times\Bigl\{\delta_{{\bf p}_1,{\bf p}}\left[
g_{ik}e^{i{\bf p}_c\cdot{\bf r}/\hbar}\Phi^*_k
\langle a^{\dagger}_{j{\bf p}_1}a_{i{\bf p}_2}a_{k{\bf p}_3}\rangle_t
-g_{kj}e^{-i{\bf p}_c\cdot{\bf r}/\hbar}
\langle a^{\dagger}_{k{\bf p}_3}a^{\dagger}_{j{\bf p}_2}
a_{i{\bf p}_3}\rangle_t\Phi_k\right] \cr
&&-\delta_{{\bf p}_1,{\bf p}}\left[g_{kj}e^{i{\bf p}_c\cdot{\bf r}/\hbar}\Phi^*_k
\langle a^{\dagger}_{j{\bf p}_1}a_{i{\bf p}_2}a_{k{\bf p}_3}\rangle_t
-g_{ik}e^{-i{\bf p}_c\cdot{\bf r}/\hbar}
\langle a^{\dagger}_{k{\bf p}_3}a^{\dagger}_{j{\bf p}_2}a_{i{\bf p}_1}\rangle_t
\Phi_k\right] \cr
&&-\delta_{{\bf p}_1,{\bf p}}\bigl[g_{kj}e^{i{\bf p}_c\cdot{\bf r}/\hbar}\Phi^*_j
\langle a^{\dagger}_{k{\bf p}_1}a_{k{\bf p}_2}a_{i{\bf p}_3}\rangle_t \cr
&&-g_{ik}e^{-i{\bf p}_c\cdot{\bf r}/\hbar}
\langle a^{\dagger}_{j{\bf p}_3}a^{\dagger}_{k{\bf p}_2}a_{k{\bf p}_1}\rangle_t
\Phi_i\bigr] \Bigr\},
\end{eqnarray}
where the three-field correlation function is given in Eq.~(\ref{3field_p}).

In order to denote the collision term in a compact form, it is useful
to introduce the three matrices:
\begin{eqnarray}
A_{ij}^{\stackrel{<}{>}}(C;{\bf p}_2,{\bf p}_3)&\equiv&\sum_{kl}
g_{kj}g_{il}\{n_{ckl}
W^{\stackrel{<}{>}}_{lk}({\bf p}_2)W^{\stackrel{<}{>}}_{ij}({\bf p}_3)\cr
&&~~+n_{ckl}W^{\stackrel{<}{>}}_{ik}({\bf p}_2)
W^{\stackrel{<}{>}}_{lj}({\bf p}_3)\},
\label{A<C1}
\end{eqnarray}
\begin{eqnarray}
A_{ij}^{\stackrel{<}{>}}({\bf p}_1;C,{\bf p}_3)&\equiv&\sum_{kl}
g_{kj}g_{il}\{W^{\stackrel{>}{<}}_{kl}({\bf p}_1)
n_{clk}W^{\stackrel{<}{>}}_{ij}({\bf p}_3)\cr
&&~~+W^{\stackrel{>}{<}}_{kl}({\bf p}_1)n_{cik}
W^{\stackrel{<}{>}}_{lj}({\bf p}_3)\},
\label{A<C2}
\end{eqnarray}
\begin{eqnarray}
A_{ij}^{\stackrel{<}{>}}({\bf p}_1;{\bf p}_2,C)&\equiv&\sum_{kl}
g_{kj}g_{il}\{W^{\stackrel{>}{<}}_{kl}({\bf p}_1)
W^{\stackrel{<}{>}}_{lk}({\bf p}_2)n_{cij}\cr
&&~~+W^{\stackrel{>}{<}}_{kl}({\bf p}_1)W^{\stackrel{<}{>}}_{ik}({\bf p}_2)
n_{clj}\},
\label{A<C3}
\end{eqnarray}
We then find
\begin{equation}
-\frac{i}{\hbar}{\rm tr}\tilde\rho(t,t_0)[\uuline{\hat W},\hat{H}'_3]
=\uuline{I_c} [\uuline{W}({\bf p})]+\frac{i}{\hbar}[\uuline{W}({\bf p}),
\delta\uuline{U}{}_c({\bf p})],
\end{equation}
where $\uuline{I}{}_c$ represents collisions between the condensate
and noncondensate atoms:
\begin{eqnarray}
&&\uuline{I}{}_c[\uuline{W}({\bf p})] \cr
&&=
\frac{\pi}{\hbar V}\sum_{{\bf p}_1,{\bf p}_2,{\bf p}_3}
\delta_{{\bf p}_c+{\bf p}_1,{\bf p}_2+{\bf p}_3}
\delta(\varepsilon_c+\tilde\varepsilon_{p_1}-\tilde\varepsilon_{p_2}
-\tilde\varepsilon_{p_3})
\cr
&&\times\Biggl\{\delta_{ {\bf p},{\bf p}_1}
\left[\left\{{\uuline{1}}+\uuline{W}({\bf p}_1),
\uuline{A}^<(C;{\bf p}_2,{\bf p}_3)\right\}
-\left\{\uuline{W}({\bf p}_1),\uuline{A}^>(C;{\bf p}_2,{\bf p}_3)\right\}\right] \cr
&&+\delta_{{\bf p},{\bf p}_2}
\left[\left\{{\uuline{1}}+\uuline{W}({\bf p}_2),
\uuline{A}^<({\bf p}_3;C,{\bf p}_1)\right\}
-\left\{\uuline{W}({\bf p}_2),\uuline{A}^>({\bf p}_3;C,{\bf p}_1)\right\}\right] \cr
&&+\delta_{{\bf p},{\bf p}_3}
\left[\left\{{\uuline{1}}+\uuline{W}({\bf p}_3),
\uuline{A}^<({\bf p}_2;{\bf p}_1,C)\right\}
-\left\{\uuline{W}({\bf p}_3),\uuline{A}^>({\bf p}_2;{\bf p}_1;C)\right\}\right]
\Biggr\}.
\label{Ic_general}
\end{eqnarray}
The second term $\delta\uuline{U}{}_c$ represents the second-order
correction to the effective coupling field:
\begin{eqnarray}
\delta\uuline{U}{}_c({\bf p})&=&\frac{1}{\hbar V}
\sum_{{\bf p}_1,{\bf p}_2,{\bf p}_3}
{\cal P}\frac{1}{\varepsilon_c+\tilde\varepsilon_{p_1}
-\tilde\varepsilon_{p_2}-\tilde\varepsilon_{p_3}}
\delta_{{\bf p}_c+{\bf p}_1,{\bf p}_2+{\bf p}_3} \cr
&&\times\Biggl\{
\delta_{{\bf p},{\bf p}_1}
\left[\uuline{A}^>(C;{\bf p}_3,{\bf p}_2)-\uuline{A}^<(C;{\bf p}_3,{\bf p}_2)\right]\cr
&&+\delta_{{\bf p},{\bf p}_2}\left[\uuline{A}^>({\bf p}_3;C,{\bf p}_1)
-\uuline{A}^<({\bf p}_3;C,{\bf p}_1)\right] \cr
&&+\delta_{{\bf p},{\bf p}_3}\left[\uuline{A}^>({\bf p}_2;{\bf p}_1;C)
-\uuline{A}^<({\bf p}_2;{\bf p}_1,C)\right]\Biggr\}.
\label{deltaUc}
\end{eqnarray}

Finally, we consider the contribution from $\hat H_4'$:
\begin{eqnarray}
&&\frac{1}{i\hbar}{\rm tr}\tilde\rho(t,t_0)
[\hat{W}_{ij}({\bf r},{\bf p},t),H'_4(t)] \cr
&&\simeq \frac{1}{i\hbar}\sum_{\bf q}e^{i{\bf q}\cdot{\bf r}/\hbar}
\Biggl(\frac{1}{V}\sum_k\sum_{{\bf p}_1,{\bf p}_2,{\bf p}_3,{\bf p}_4}
\delta_{{\bf p}_1+{\bf p}_2,{\bf p}_3+{\bf p}_4} \cr
&&\times [\delta_{{\bf p}_1,{\bf p}+{\bf q}/2}g_{ik}
\langle a^{\dagger}_{j{\bf p}-{\bf q}/2}a^{\dagger}_{k{\bf p}_2}
a_{k{\bf p}_3}a_{i{\bf p}_4}\rangle_t \cr
&&-\delta_{{\bf p}_4,{\bf p}+{\bf q}/2}g_{kj}
\langle a^{\dagger}_{j{\bf p}_1}a^{\dagger}_{k{\bf p}_2}
a_{k{\bf p}_3}a_{i{\bf p}+{\bf q}/2}\rangle_t]\Biggr) \cr 
&&-\sum_k\bigl\{g_{kj}[\tilde n_{kj}W_{ik}({\bf p})+
\tilde n_{kk}W_{ij}({\bf p})] \cr
&&-g_{ik}[\tilde n_{ik}W_{kj}({\bf p})
+\tilde n_{kk}W_{ij}({\bf p})]\bigr\}.
\label{H4_terms}
\end{eqnarray}
We now need to evaluate the four-field correlation function:
\begin{eqnarray}
&&\langle 
a^{\dagger}_{j{\bf p}_1}a^{\dagger}_{k{\bf p}_2}
a_{k{\bf p}_3}a_{i{\bf p}_4}
\rangle_t \cr
&&={\rm tr}\hat\rho(t_0)\Bigl\{\hat\EuScript{U}_0^{\dagger}(t,t_0)
a^{\dagger}_{j{\bf p}_1}a^{\dagger}_{k{\bf p}_2}
a_{k{\bf p}_3}a_{i{\bf p}_4}
\hat\EuScript{U}_0(t,t_0) \cr
&&-\frac{i}{\hbar}\int_{t_0}^tdt'
e^{i(\tilde\varepsilon_{p_1}+\tilde\varepsilon_{p_2}-
\tilde\varepsilon_{p_3}-\tilde\varepsilon_{p_4})} \cr
&&\times \hat\EuScript{U}_0^{\dagger}(t',t_0)
[a^{\dagger}_{j{\bf p}_1}a^{\dagger}_{k{\bf p}_2}
a_{k{\bf p}_3}a_{i{\bf p}_4}, \hat H'_4(t')]
\hat\EuScript{U}_0(t',t_0)\Bigr\}.
\label{4_field}
\end{eqnarray}
We find that the first term of order $g^0$ makes no contribution
to Eq.~(\ref{H4_terms}).
Using Wick's theorem to evaluate the second order term, we also
find that mean field contribution in the last two terms of Eq.~(\ref{H4_terms})
are canceled out with several terms in Eq.~(\ref{4_field}).
The relevant terms are often referred to as ``connected diagrams".
We thus obtain the relevant contribution:
\begin{eqnarray}
&&\langle
a^{\dagger}_{j{\bf p}_1}a^{\dagger}_{k{\bf p}_2}
a_{k{\bf p}_3}a_{i{\bf p}_4} \rangle_t \cr
&&\simeq
-\frac{i}{V}\left[\pi\delta(\tilde\varepsilon_{p_1}+\tilde\varepsilon_{p_2}
-\tilde\varepsilon_{p_3}-\tilde\varepsilon_{p_4})
+i{\cal P}\frac{1}{\tilde\varepsilon_{p_1}+\tilde\varepsilon_{p_2}
-\tilde\varepsilon_{p_3}-\tilde\varepsilon_{p_4}}\right] \cr
&&\times\delta_{{\bf p}_1+{\bf p}_2,{\bf p}_3+{\bf p}_4}\sum_{mn}g_{mn}
 \Bigl\{
   W_{mj}({\bf p}_1)W_{nk}({\bf p}_2)[\delta_{kn}+W_{kn}({\bf p}_3)]
[\delta_{im}+W_{im}({\bf p}_4)] \cr
&&+W_{mj}({\bf p}_1)W_{nk}({\bf p}_2)[\delta_{km}+W_{km}({\bf p}_3)]
[\delta_{in}+W_{in}({\bf p}_4)] \cr
&&-[\delta_{mj}+W_{mj}({\bf p}_1)][\delta_{nk}+W_{nk}({\bf p}_2)]
W_{kn}({\bf p}_3)W_{im}({\bf p}_4) \cr
&&-[\delta_{mj}+W_{mj}({\bf p}_1)][\delta_{nk}+W_{nk}({\bf p}_2)]
W_{km}({\bf p}_3)W_{in}({\bf p}_4)\Bigr\}.
\end{eqnarray}
Using the matrices defined above, we can write this $\hat{H}'_4$ contribution
to the kinetic equation as
\begin{equation}
\frac{1}{i\hbar}{\rm tr}\tilde\rho(t,t_0)
[\uuline{\hat{W}}({\bf r},{\bf p},t),\hat{H}'_4(t)]=
\uuline{I}{}_n[\uuline{W}({\bf p})]
+\frac{i}{\hbar}[\uuline{W}({\bf p}),\delta\uuline{U}{}_n({\bf p})],
\end{equation}
where $\uuline{I}{}_n$ represents the collisions between noncondensate atoms
\begin{eqnarray}
\uuline{I}{}_n[\uuline{W}({\bf p})]
&=&
\frac{\pi}{\hbar V^2}\sum_{{\bf p}_2,{\bf p}_3,{\bf p}_4}
\delta(\tilde\varepsilon_p+\tilde\varepsilon_{p_2}-\tilde\varepsilon_{p_3}
-\tilde\varepsilon_{p_4}))
\delta_{{\bf p}+{\bf p}_2,{\bf p}_3+{\bf p}_4} \cr
&&\times
\Bigl[\left\{{\uuline{1}}+\uuline{W}({\bf p}),
\uuline{A}^<({\bf p}_2;{\bf p}_3,{\bf p}_4)\right\} \cr
&&-\left\{\uuline{W}({\bf p}),\uuline{A}^>({\bf p}_2;{\bf p}_3,{\bf p}_4)\right\}\Bigr],
\label{In}
\end{eqnarray}
\begin{eqnarray}
\delta\uuline{U}{}_n({\bf p})&=&\frac{1}{\hbar V^2}
\sum_{{\bf p}_2,{\bf p}_3,{\bf p}_4}
{\cal P}\frac{1}{\tilde\varepsilon_p+\tilde\varepsilon_{p_2}
-\tilde\varepsilon_{p_3}-\tilde\varepsilon_{p_4}}
\delta_{{\bf p}+{\bf p}_2,{\bf p}_3-{\bf p}_4} \cr
&&\times
\left[\uuline{A}^>({\bf p}_2;{\bf p}_3,{\bf p}_4)
-\uuline{A}^<({\bf p}_2;{\bf p}_3,{\bf p}_4)\right]
\label{deltaUn}
\end{eqnarray}

We now comment on the spin rotation terms involving $\delta \uuline{U}{}_n$ and
$\delta \uuline{U}{}_c$.
This type of contribution is known as the ``off-energy-shell" term.
In the high-temperature (or low-density) Maxwell-Boltzmann limit
$T\gg T_c$, one can show that
$\delta\uuline{U}({\bf p})$ is canceled out by a term second-order
in the $t$-matrix in the mean-field term \cite{Jeon1988a,Ruckenstein1985a}.
In the degenerate quantum gas, the role of the off-energy-shell term is
not obvious, and there are several discussions in literature
(see discussions in Jeon and Mullin~\cite{Jeon1988a}).
In any event, in our present treatment we do not include renormalization
effects consistently to order $g^2$. 
We thus drop the off-energy-shell contribution to the streaming term.
For consistency, we also drop the anomalous correlation function
and the principal-value part of $\uuline{R}$ in the GP equation.
This means that we drop the second-order correction to the condensate chemical potential.
Including these second-order contribution will require more careful treatment of the
many-body perturbation theory.

Finally, we can replace the momentum sum $1/V\sum_{\bf p}$ by the integral
$\int d{\bf p}/(2\pi\hbar)^3$ and the Kronecker delta function 
$V\delta_{{\bf p},{\bf p}'}$ by the Dirac delta function $(2\pi\hbar)^3
\delta({\bf p}-{\bf p}')$.
We then obtain
\begin{eqnarray}
&&\uuline{I}{}_n[\uuline{W}({\bf p})]=
\frac{\pi}{\hbar}\int\frac{d{\bf p}_2}{(2\pi\hbar)^3}
\int\frac{d{\bf p}_3}{(2\pi\hbar)^3}\int d{\bf p}_4 \cr
&&\times\delta(\tilde\varepsilon_p+\tilde\varepsilon_{p_2}-\tilde\varepsilon_{p_3}
-\tilde\varepsilon_{p_4}))
\delta({\bf p}+{\bf p}_2-{\bf p}_3-{\bf p}_4) \cr
&&\times
\left[\left\{{\uuline{1}}+\uuline{W}({\bf p}),\uuline{A}^<({\bf p}_2;{\bf p}_3,{\bf p}_4)\right\}
-\left\{\uuline{W}({\bf p}),\uuline{A}^>({\bf p}_2;{\bf p}_3,{\bf p}_4)\right\}\right],
\label{In_general2}
\end{eqnarray}
\begin{eqnarray}
&&\uuline{I}{}_c[\uuline{W}({\bf p})]=
\frac{\pi}{\hbar}\int\frac{d{\bf p}_1}{(2\pi\hbar)^3}
\int d{\bf p}_2 \int d{\bf p}_3 \cr
&&\times\delta(\varepsilon_c+\tilde\varepsilon_{p_1}-\tilde\varepsilon_{p_2}
-\tilde\varepsilon_{p_3}))
\delta(m{\bf v}_c+{\bf p}_1-{\bf p}_2-{\bf p}_3) \cr
&&\times\{\delta({\bf p}-{\bf p}_1)
\left[\left\{{\uuline{1}}+\uuline{W}({\bf p}_1),\uuline{A}^<(C;{\bf p}_2,{\bf p}_3)\right\}
-\left\{\uuline{W}({\bf p}_1),\uuline{A}^>(C;{\bf p}_2,{\bf p}_3)\right\}\right] \cr
&&+\delta({\bf p}-{\bf p}_2)
\left[\left\{{\uuline{1}}+\uuline{W}({\bf p}_2),\uuline{A}^<({\bf p}_3;C,{\bf p}_1)\right\}
-\left\{\uuline{W}({\bf p}_2),\uuline{A}^>({\bf p}_3;C,{\bf p}_1)\right\}\right] \cr
&&+\delta({\bf p}-{\bf p}_3)
\left[\left\{{\uuline{1}}+\uuline{W}({\bf p}_3),\uuline{A}^<({\bf p}_2;{\bf p}_1,C)\right\}
-\left\{\uuline{W}({\bf p}_3),\uuline{A}^>({\bf p}_2;{\bf p}_1;C)\right\}\right].
\label{Ic_general2}
\end{eqnarray}
\begin{eqnarray}
\uuline{R}&=&\pi\int\frac{d{\bf p}_1}{(2\pi\hbar)^3}
\int\frac{d{\bf p}_2}{(2\pi\hbar)^3}\int d{\bf p}_3 \cr
&&\times\delta(m{\bf v}_c+{\bf p}_1-{\bf p}_2-{\bf p}_3)
\delta(\varepsilon_c+\tilde\varepsilon_{p_1}-
\tilde\varepsilon_{p_2}-\tilde\varepsilon_{p_3}) \cr
&&[\uuline{A}^>({\bf p}_1;{\bf p}_2,{\bf p}_3)-
\uuline{A}^<({\bf p}_1;{\bf p}_2,{\bf p}_3)],
\label{R_general2}
\end{eqnarray}
where the various $A_{ij}^{\stackrel{<}{>}}$ are given in Eq.~(\ref{A<}) and
Eqs.~(\ref{A<C1})-(\ref{A<C3}).

\section{Spin transport relaxation times}
\label{linear_collision}

In this Appendix, we calculate various spin transport relaxation times that appear
in the spin hydrodynamic equations.
We first consider the following relaxation term:
\begin{eqnarray}
\left. \frac{\partial \vec{S}_n }{\partial t}\right|_{\rm coll}
&=&\rm{Tr}(\uuline{\vec{\sigma}}\,\uuline{\Gamma}{}_c[\uuline{W}^{\rm leq}] \cr
&=&-\frac{\pi g^2 n_c}{\hbar} \int\frac{d{\bf p}_1}{(2\pi\hbar)^3}
\int \frac{d{\bf p}_2}{(2\pi\hbar)^3}\int d{\bf p}_3 \cr
&&\times\delta(m{\bf v}_c+{\bf p}_1-{\bf p}_2-{\bf p}_3)
\delta(\varepsilon_c+\tilde\varepsilon_{p_1}-
\tilde\varepsilon_{p_2}-\tilde\varepsilon_{p_3}) \cr
&&\times\sum_{s,s'=\pm1}[1-e^{-\beta(\tilde\mu_{s'}-\mu_c)}] \cr
&&\times(\vec{e}_c+s'\vec{e}_n)(1+\delta_{s,s'})(1+f_{1s})f_{2s}f_{3s'}.
\label{Sn_collWleq}
\end{eqnarray}
We then linearize Eq.~(\ref{Sn_collWleq}) in $\mu_{\rm diff}$ and 
$\vec{e}_n-\vec{e}_c$ to find
\begin{eqnarray}
\left.\frac{\partial\vec{S}_n}{\partial t}\right|_{\rm coll}
=-\frac{\tilde n\beta\mu_{\rm diff}}{\tilde\tau_c^{\parallel}}
\left(\frac{\vec{e}_n+\vec{e}_c}{2}\right)
-\frac{\tilde n}{\tilde\tau_c^{\perp}}(\vec{e}_n-\vec{e}_c),
\label{Sn_coll3}
\end{eqnarray}
where the two relaxation times are given by
\begin{eqnarray}
\frac{1}{\tilde\tau_c^{\parallel}}=
\frac{n_c}{\tilde n}\frac{1}{\tau_c^{\parallel}}
&=&\frac{2\pi g^2n_c}{\hbar\tilde n}
\int\frac{d{\bf p}_1}{(2\pi\hbar)^3}\int\frac{d{\bf p}_2}{(2\pi\hbar)^3}
\int d{\bf p}_3 \cr
&&\times\delta({\bf p}_1-{\bf p}_2-{\bf p}_3) 
\delta(\mu_c+{\bf p}_1-{\bf p}_2-{\bf p}_3)\cr
&&\times\sum_s(1+\delta_{s,+})(1+f_{1s})f_{2s}f_{3+},
\end{eqnarray}
\begin{eqnarray}
\frac{1}{\tilde\tau_c^{\perp}}&=&\frac{n_c}{\tilde n}\frac{1}{\tau_c^{\perp}} 
=\frac{\pi g^2n_c}{\hbar\tilde n}
\int\frac{d{\bf p}_1}{(2\pi\hbar)^3}\int\frac{d{\bf p}_2}{(2\pi\hbar)^3}
\int d{\bf p}_3 \cr
&&\times \delta({\bf p}_1-{\bf p}_2-{\bf p}_3)
\delta(\mu_c+{\bf p}_1-{\bf p}_2-{\bf p}_3)\cr
&&\times\sum_s(1+\delta_{s,-})[f_{1s}(1+f_{2s})(1+f_{3-})
-(1+f_{1s})f_{2s}f_{3-}].
\end{eqnarray}

We next calculate the relaxation times associated with the spin current.
Linearized collision integrals are given by
\begin{equation}
\delta\uuline{I}{}_n=\delta\uuline{I}{}_n^{\parallel}+
\delta\uuline{I}{}_n^{\perp},
\end{equation}
\begin{equation}
\delta\uuline{I}{}_c=\delta\uuline{I}{}_c^{\parallel}+
\delta\uuline{I}{}_c^{\perp},
\end{equation}
where
\begin{eqnarray}
\delta\uuline{I}{}_n^{\parallel}&=&\frac{\pi g^2}{2\hbar}\int\frac{d{\bf p}_1}{(2\pi\hbar)^3}
\int\frac{d{\bf p}_2}{(2\pi\hbar)^3}\int d{\bf p}_3\int d{\bf p}_4 \cr
&&\times \delta({\bf p}_1+{\bf p}_2-{\bf p}_3-{\bf p}_4)\delta(\tilde\varepsilon_{p_1}
+\tilde\varepsilon_{p_2}-\tilde\varepsilon_{p_3}-\tilde\varepsilon_{p_4}) \cr
&&\times[\delta({\bf p}-\delta{\bf p}_1)-\delta({\bf p}-{\bf p}_4)] \cr
&&\times\sum_{s,s'}(\uuline{1}+s\vec{e}_n\cdot\vec{\uuline{\sigma}})(1+\delta_{s,s'})
(1+f_{1s})(1+f_{2s'})f_{3s'}f_{4s} \cr
&&\times(\psi_{4s}^{\parallel}+\psi_{3s'}^{\parallel}-\psi_{2s'}^{\parallel}
-\psi_{1s}^{\parallel}),
\end{eqnarray}
\begin{eqnarray}
\delta\uuline{I}{}_n^{\perp}&=&\frac{g^2\pi}{4\hbar}\int\frac{d{\bf p}_1}{(2\pi\hbar)^3}
\int\frac{d{\bf p}_2}{(2\pi\hbar)^3}\int d{\bf p}_3 \int d{\bf p}_4 \cr
&&\times \delta({\bf p}_1+{\bf p}_2-{\bf p}_3-{\bf p}_4)\delta(\tilde\varepsilon_{p_1}
+\tilde\varepsilon_{p_2}-\tilde\varepsilon_{p_3}-\tilde\varepsilon_{p_4}) \cr
&&\times[\delta({\bf p}-{\bf p}_1)-\delta({\bf p}-{\bf p}_4)] \cr
&&\times \sum_{s,s'}
[(1+f_{1s})(1+f_{2s'})f_{3s'}f_{4-s}-f_{1s}f_{2s'}(1+f_{3s'})(1+f_{4-s})] \cr
&&\times s\vec{\uuline{\sigma}}\cdot[\vec{\psi}^{\perp}_1-\vec{\psi}^{\perp}_4
+\delta_{s,s'}(\vec{\psi}^{\perp}_2-\vec{\psi}^{\perp}_4)
+\delta_{s,-s'}(\vec{\psi}^{\perp}_1-\vec{\psi}^{\perp}_3)],
\end{eqnarray}
\begin{eqnarray}
\delta\uuline{I}{}_c^{\parallel}&=&\frac{\pi g^2 n_c}{\hbar}
\int\frac{d{\bf p}_1}{(2\pi\hbar)^3}\int d{\bf p}_2 \int d{\bf p}_3 \cr
&&\times \delta({\bf p}_1-{\bf p}_2-{\bf p}_3)\delta(\mu_c+\tilde\varepsilon_{p_1}
-\tilde\varepsilon_{p_2}-\tilde\varepsilon_{p_3}) \cr
&&\times\sum_s([\delta({\bf p}-{\bf p}_1)-\delta({\bf p}-{\bf p}_2)-
\delta({\bf p}-{\bf p}_3)]\uuline{1} \cr
&&+\{s[\delta({\bf p}-{\bf p}_1)-\delta({\bf p}-{\bf p}_2)]-
\delta({\bf p}-{\bf p}_3)\}\vec{e}_n\cdot\vec{\uuline{\sigma}})\cr
&&\times(1+\delta_{s,+})(1+f_{1s})f_{2s}f_{3+}(\psi^{\parallel}_{3+}
+\psi^{\parallel}_{2s}-\psi^{\parallel}_{1s}),
\end{eqnarray}
\begin{eqnarray}
\delta\uuline{I}{}_c^{\perp}&=&\frac{\pi g^2 n_c}{2\hbar}
\int\frac{d{\bf p}_1}{(2\pi\hbar)^3}\int d{\bf p}_2 \int d{\bf p}_3 \cr
&&\times \delta({\bf p}_1-{\bf p}_2-{\bf p}_3)\delta(\mu_c+\tilde\varepsilon_{p_1}
-\tilde\varepsilon_{p_2}-\tilde\varepsilon_{p_3}) \cr
&&\times\sum_s([\delta({\bf p}-{\bf p}_1)-\delta({\bf p}-{\bf p}_2)] \cr
&&\times [(1+f_{1s})f_{2-s}f_{3+}-f_{1s}(1+f_{2-s})(1+f_{3+})]\cr
&&\times[s(\vec{\psi}^{\perp}_1-\vec{\psi}^{\perp}_2)
-\delta_{s,-}(\vec{\psi}^{\perp}_1-\vec{\psi}^{\perp}_3)
-\delta_{s,+}\vec{\psi}^{\perp}_2]\cdot\vec{\uuline{\sigma}}\cr
&&-\delta({\bf p}-{\bf p}_3)
[(1+f_{1s})f_{2s}f_{3-}-f_{1s}(1+f_{2s})(1+f_{3-})]\cr
&&\times[-\vec{\psi}^{\perp}_3+\delta_{s,+}(\vec{\psi}^{\perp}_1-\vec{\psi}^{\perp}_3)
-\delta_{s,-}\vec{\psi}_2^{\perp}]\cdot\vec{\uuline{\sigma}}).
\end{eqnarray}
When linearizing $\uuline{I}{}_c$, we assumed $\vec{e}_n=\vec{e}_c$ and
$\mu_{\rm diff}=0$.

Taking moments of these linearized collision integrals with using Eq.~(\ref{psi_correction}), we find
\begin{equation}
\int \frac{d{\bf p}}{(2\pi\hbar)^3}{\bf p}
{\rm Tr}\left(\uuline{\vec{\sigma}}\delta\uuline{I}{}_n^{\parallel}\right)
=-\frac{\vec{\bf J}_n^{\parallel}}{\tau_{D,n}^{\parallel}},~~
\int \frac{d{\bf p}}{(2\pi\hbar)^3}{\bf p}
{\rm Tr}\left(\uuline{\vec{\sigma}}\delta\uuline{I}{}_n^{\perp}\right)
=-\frac{\vec{\bf J}_n^{\perp}}{\tau_{D,n}^{\perp}},
\end{equation}
and
\begin{equation}
\int \frac{d{\bf p}}{(2\pi\hbar)^3}{\bf p}
{\rm Tr}\left(\uuline{\vec{\sigma}}\delta\uuline{I}{}_c^{\parallel}\right)
=-\frac{\vec{\bf J}_n^{\parallel}}{\tau_{D,c}^{\parallel}},~~
\int \frac{d{\bf p}}{(2\pi\hbar)^3}{\bf p}
{\rm Tr}\left(\uuline{\vec{\sigma}}\delta\uuline{I}{}_c^{\perp}\right)
=-\frac{\vec{\bf J}_n^{\perp}}{\tau_{D,c}^{\perp}},
\end{equation}
where the four relaxation times are given by
\begin{eqnarray}
\frac{1}{\tau_{D,n}^{\parallel}}&=&\frac{\pi g^2}{3\hbar}\frac{\tilde n}
{\tilde{n}_+\tilde{n}_-}
\frac{\beta}{m}\int\frac{d{\bf p}_1}{(2\pi\hbar)^3}
\int\frac{d{\bf p}_2}{(2\pi\hbar)^3}\int\frac{d{\bf p}_3}{(2\pi\hbar)^3}
\int d{\bf p}_4 \cr
&&\times\delta({\bf p}_1+{\bf p}_2-{\bf p}_3-{\bf p}_4)
\delta(\tilde\varepsilon_{p_1}+\tilde\varepsilon_{p_2}
-\tilde\varepsilon_{p_3}-\tilde\varepsilon_{p_4})\cr
&&\times({\bf p}_1-{\bf p}_4)^2(1+f_{1+})(1+f_{2-})f_{3-}f_{4+},
\end{eqnarray}
\begin{eqnarray}
\frac{1}{\tau_{D,c}^{\parallel}}&=&\frac{2\pi g^2n_c}{3\hbar}\frac{\tilde n}
{\tilde {n}_+\tilde {n}_-}
\frac{\beta}{m}\int\frac{d{\bf p}_1}{(2\pi\hbar)^3}
\int\frac{d{\bf p}_2}{(2\pi\hbar)^3}\int d{\bf p}_3 \cr
&&\times
\delta({\bf p}_1-{\bf p}_2-{\bf p}_3)
\delta(\mu_c+\tilde\varepsilon_{p_1}-\tilde\varepsilon_{p_2}
-\tilde\varepsilon_{p_3})\cr
&&\times p_3^2(1+f_{1-})f_{2-}f_{3+},
\end{eqnarray}
\begin{eqnarray}
\frac{1}{\tau_{D,n}^{\perp}}&=&\frac{\pi g^2}{6\hbar m\sum_s s\tilde P_s}
\int\frac{d{\bf p}_1}{(2\pi\hbar)^3}\int\frac{d{\bf p}_2}{(2\pi\hbar)^3}
\int\frac{d{\bf p}_3}{(2\pi\hbar)^3}\int d{\bf p}_4 \cr
&&\times\delta({\bf p}_1+{\bf p}_2-{\bf p}_3-{\bf p}_4)
\delta(\tilde\varepsilon_{p_1}+\tilde\varepsilon_{p_2}
-\tilde\varepsilon_{p_3}-\tilde\varepsilon_{p_4}) \cr
&&\times\sum_{s,s'}[(1+f_{1-s})(1+f_{2s'})f_{3s'}f_{4s}
-f_{1-s}f_{2s'}(1+f_{3s'})(1+f_{4s})]\cr
&&\times[s({\bf p}_1-{\bf p}_4)^2+s'({\bf p}_1-{\bf p}_4)\cdot({\bf p}_1-{\bf p}_3)],
\end{eqnarray}
\begin{eqnarray}
\frac{1}{\tau_{D,c}^{\perp}}&=&\frac{\pi g^2}{3\hbar m\sum_s s\tilde P_s}
\int\frac{d{\bf p}_1}{(2\pi\hbar)^3}\int\frac{d{\bf p}_2}{(2\pi\hbar)^3}
\int d{\bf p}_3 \cr
&&\times\delta({\bf p}_1-{\bf p}_2-{\bf p}_3)
\delta(\mu_c+\tilde\varepsilon_{p_1}-\tilde\varepsilon_{p_2}
-\tilde\varepsilon_{p_3})\cr
&&\times\sum_{s}\{
s[(1+f_{1-s})f_{2s})f_{3+}-f_{1-s}(1+f_{2s})(1+f_{3+})]\cr
&&-[(1+f_{1-s})f_{2-s}f_{3-}-f_{1-s}(1+f_{2-s})(1+f_{3-})] \}
\cr
&&\times(p_3^2+s{\bf p}_2\cdot{\bf p}_3)\}.
\end{eqnarray}
Finally, the longitudinal spin diffusion time $\tau_D^{\parallel}$ is given by
\begin{equation}
\frac{1}{\tau_D^{\parallel}}=\frac{1}{\tau_{D,n}^{\parallel}}+
\frac{1}{\tau_{D,c}^{\parallel}},
\end{equation}
and the transverse diffusion relaxation time is given by
\begin{equation}
\frac{1}{\tau_D^{\perp}}=\frac{1}{\tau_{D,n}^{\perp}}+
\frac{1}{\tau_{D,c}^{\perp}}.
\end{equation}

\end{appendix}



\end{document}